\documentclass[prb,twocolumn,showpacs,showkeys,superscriptaddress]{revtex4}
\usepackage{latexsym}
\usepackage{graphicx}
\usepackage{dcolumn}
\usepackage{amsmath}
\usepackage{bm}
\usepackage{epstopdf}

\begin{document}

\title{Experimental study of the competition between Kondo and RKKY interactions for Mn spins in 
a model alloy system}

\author{J.J.~Pr\'ejean}
\email[Corresponding author: ]{jean-jacques.prejean@grenoble.cnrs.fr}

\author{E. Lhotel}

\author{A. Sulpice}

\affiliation{ Centre de Recherches sur les Tr\`es Basses
Temp\'eratures, CNRS, BP 166\\
F-38042 Grenoble Cedex-9, France}

\author{F. Hippert}
\affiliation{Laboratoire des Mat\'eriaux et du G\'enie Physique, ENSPG, BP46, 
38402 Saint Martin d'H\`eres Cedex, France.}

\begin{abstract}
The quasicrystal $Al-Pd-Mn$ is a model system for an experimental
study of the competition between Ruderman-Kittel-Kasuya-Yoshida (RKKY) 
and Kondo interactions. First, specific of such alloys, only a few $Mn$ atoms carry an 
effective spin and their concentration $x$ is tunable over several orders of magnitude, even though the $Mn$ amount is almost constant. Second, the characteristic energy scales for the interactions lie in the Kelvin range. Hence we could study the magnetization on both side of these energy scales, covering a range of temperatures [0.1-100 K] and magnetic fields ($\mu_B H/k_B=$ 0  to 5 K) for 22 samples and $x$ varying over 2 decades. Using very general Kondo physics arguments, and thus carrying out the data analysis with no preconceived model, we found a very robust and simple result: The magnetization is a sum of a pure Kondo ($T_K=3.35K$) and a pure RKKY contributions, whatever the moment concentration is and this surprisingly  up to the concentration where the RKKY couplings dominate fully and thus cannot be considered as a perturbation. 
\end{abstract}

\maketitle


\section{Introduction}

\label{intro}

In the early seventies, experimentalists met the problem of the competition between Kondo and 
RKKY interactions when studying the magnetic properties of dilute alloys. The latter were noble metals in which one adds $3d$ 
transition atoms, called "impurities", such as $Fe$ in $Cu$\cite{Tholence}. 
Later, this problem arose again when concentrated alloys, containing rare earth 
elements such as $Ce$ or $Yb$, became the subject of numerous 
studies. From a theoretical point of view, to 
solve this problem remains a challenge in contrast with the two 
limiting cases (pure RKKY, pure Kondo) which are well understood. It 
is mainly because the two interactions originate from the same 
anti-ferromagnetic coupling $\lambda {\bf
S.s}$ which develops on the moment 
site between the spin $S$ of the local moment and the spins $s$ of the 
conduction electrons. Consequently the same ingredients govern 
the two characteristic energy scales which emerge in this problem, 
although in a different way. These ingredients are the exchange 
constant $\lambda $ and the electronic density of states (DOS) at the Fermi 
level, denoted $\rho$ in the following. One energy scale is the \textit{bare} Kondo temperature\cite{Hewson} defined in the absence of RKKY couplings: 
$T_{K0} \propto E_{F}\times$ exp$\,\lbrack -1/\vert \lambda \vert \,\rho 
\rbrack$. It characterizes a single moment effect: The 
coupling between the local moment and the surrounding electronic 
cloud leads to the formation of a non-magnetic 
singlet when the temperature is decreased below $T_{K0}$. The 
other energy scale originates in the RKKY interaction 
$J_{ij}{\bf S_{i}S_{j}}$ which couples the spins
of two moments $i$ and $j$, $r_{ij}$ apart. This interaction is 
mediated by the spin polarization of the conduction electrons which is induced by the $\lambda {\bf
S.s}$ couplings. The $J_{ij}\propto \lambda ^{2}\rho
\times cos(2k_{F}r_{ij})/r_{ij}^{3}$ alternates in sign rapidly with 
distance, generating either a ferromagnetic or an anti-ferromagnetic 
interaction between the two moments, according to the value of $r_{ij}$. 
Thus, for given $\lambda $ and $\rho$ values, the only experimental parameter which 
can modify the relative weight of the RKKY and Kondo 
interactions is the distance $r_{ij}$ which governs the pair interaction $J_{ij}$. 
In real alloys, one deals with a collection of moments. For this case, one has first to turn to a macroscopic level to define an energy scale $T_J$ characterizing the presence of RKKY couplings.
In the same way as we defined $T_{K0}$ in the absence of RKKY coupling, we can define symmetrically $T_J$ in case of a vanishing Kondo interaction for all spins. In real alloys, when Kondo effects are negligible, one can identify the $T_J$ with the temperature at which a magnetic transition occurs due to the RKKY couplings. This transition can be either 
(anti)ferromagnetic-like for regular systems or spin-glass-like for sufficiently disordered configurations. 
Let us consider in the latter category systems where the moment concentration $x$ can be varied,  whereas the non magnetic matrix remains unchanged. Then, the mean distance $\bar{r}$ between moments is the only variable parameter which allows to change $T_J$, the $T_{K0}$ remaining constant (at least when $\lambda$ and $\rho$ are unaffected by the introduction of foreign chemical species in the matrix). In this case, $T_J$ scales with $x$ because  $J_{ij}\propto 1/r_{ij}^{3}$ and $1/ \bar{r}^{3}\propto x$. In the present paper, we are interested in this type of systems. At this stage, let us consider the optimum conditions necessary to carry out a thorough experimental study. First, the best 
is to have compounds where $T_J$, and thus $x$, can be actually varied \textit{by orders of magnitude}, because of the logarithmic Kondo behavior. 
The second condition concerns the magnitudes of the characteristic temperatures. Indeed, for each  sample, it is desirable to perform measurements at temperatures $T$ (and $\mu_{B}H/k_{B}$ for field studies) ranging from well below up to well above $T_J$ and $T_{K0}$. In practice, it means 
characteristic temperatures lying in the Kelvin range for magnetization studies because of the temperature and field ranges of the available magnetometers. 
Note that the most favorable case\cite{Noz-Blandin} is that where the magnetic 
atoms are $Mn^{++}$ with their $g=2,S= 5/2$ ground state. They exhibit a zero 
orbital moment, thus an isotropic low temperature Kondo singlet 
which is not sensitive to crystal field effects. 

\begin{figure}
\includegraphics[width=8cm]{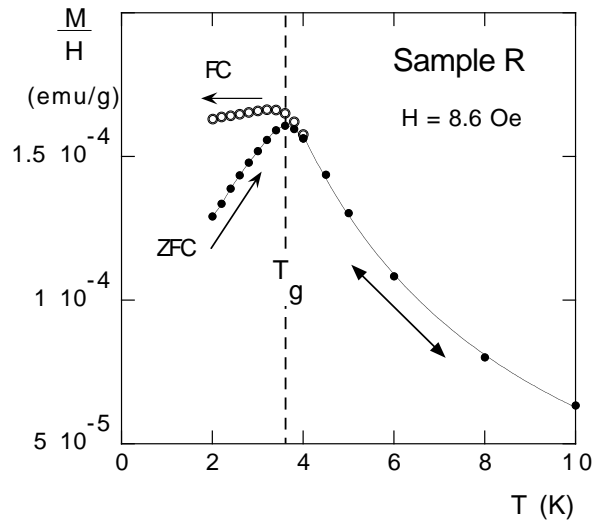}
\caption{Magnetization $M$ for $H=8.6$ Oe of sample $R$ represented in a diagram $M/H$ 
vs. $T$ for $T$ from 2K up to 10K. The zero-field-cooled (ZFC) magnetization (solid circles) exhibits a 
cusp at a temperature $T_{g}=3.6$ K. Below $T_{g}$, the field-
cooled (FC) magnetization (open circles) exhibits a plateau characteristic 
of spin-glasses. The curves are guides for the eye.} 
\label{fig00}
\end{figure}

\begin{figure}
\centering{\includegraphics[width=8cm]{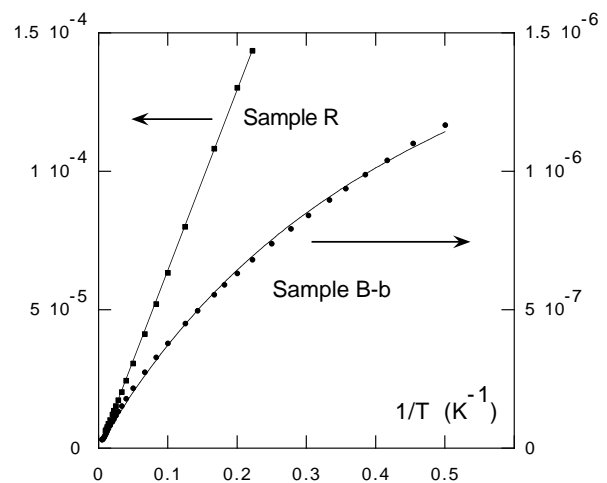}}
\caption{Susceptibility of samples $R$ and $B-b$ 
shown in a diagram $\chi$ vs. $1/T$. The solid  
curves are fits: linear fit for sample $R$ and Curie-Weiss fit 
for sample $B-b$.}
\label{fig01}
\end{figure}

To satisfy fully all the previously mentioned conditions is difficult to achieve for most dilute alloys because of the solubility limit ($Fe$ in $Cu$ for instance), the magnitude of $T_{K0}$ ($>10$ K for $Fe$ in $Cu$, $<5$ mK for $Mn$ in $Cu$) and/or a large orbital moment. In contrast, the $Al-Pd-Mn$ quasicrystalline (QC) phase satisfies these conditions. Consider two opposite behaviors described previously in separate papers. The first one is that of a $Al-Pd-Mn$ QC sample, here denoted by $R$, which behavesñ as a 
pure spin-glass\cite{art_lasjaunias}. See Fig.\ref{fig00}: Here, 
the reported magnetization $M(T)$ is measured 
in a very small d.c. field $H=8.6$ Oe. So one can identify the linear susceptibility $\chi$ with $M/H$. The sample was first cooled down to 2 K in zero field. Then the field was applied and $M$ 
was measured at increasing temperatures. In this so-called zero-field 
cooling process (ZFC), $M$ is found to exhibit a cusp at a temperature 
$T_{g}$ equal to 3.6 K. When the sample is cooled from high temperatures down to 2 K in $8.6$ Oe
(field-cooling process), the magnetization equals that measured in ZFC down to about $T_g$ but exhibits a plateau below $T_{g}$. In addition, one notes that the $\chi (T)$ follows a pure 
Curie behavior above $T_{g}$, from 4.5 K up to 100 K. This is 
clearly evident in Fig.\ref{fig01} where $\chi$ is observed 
to vary linearly with $1/T$. Altogether the susceptibility cusp, this type of observed magnetic hysteresis and the Curie behavior just above $T_{g}$ are characteristic of spin glasses. Moreover, the pure $1/T$ dependence of $\chi(T)$ from about $T_g$ up to $30\times T_g$ indicates an equal weight of the 
ferromagnetic and anti-ferromagnetic interactions. In summary, no indication of the presence of 
Kondo effects is apparent in this sample and, in this case, $T_J$ can be identified with $T_{g}$: $T_J=3.6$ K. 
The opposite case is provided by a sample, here denoted by $B-b$ as in 
Ref.\onlinecite{art_Hippert-2003}, the magnetic behavior of which could be 
dominated by the Kondo couplings. This sample is less magnetic by two orders 
of magnitudes than sample $R$: It can be readily seen from the comparison 
between the two vertical 
scales used in Fig.\ref{fig01} to represent the susceptibility of both samples in the same diagram. It follows that, in the absence of Kondo 
couplings, one should find a $T_J$ for 
sample $B-b$ about 100 times smaller than that of sample $R$, 
thus equal to 30-50 mK. Besides that, we could not detect any susceptibility cusp 
nor magnetic hysteresis down to 100 mK for this sample, 
as shown in Sec. \ref{M,X}. Another strong difference with the behavior 
of sample $R$ is the occurrence of a marked continuous curvature of 
the $\chi (1/T)$ curve over the whole [2 K-150 K] $T$-range: See Fig.\ref{fig01}. One 
notes that this curvature is negative: When $T$ is decreased, the $\chi$ is smaller and smaller than that given by $\chi \propto 1/T$ of a noninteracting ($T_{K0}=0;T_J=0$) paramagnet. This suggests the presence of 
antiferromagnetic-like interactions. The problem is to know their origin: Kondo or RKKY couplings? In particular, local chemical order which can exist in dilute alloys is known to lead to large deviations from a Curie behavior\cite{Binder} at $T\gg T_{g}$. Actually, in the present case, transport results are in strong favor of the presence of Kondo couplings: See Ref.\onlinecite{Sig-pre} where the present sample is denoted $B-II$. Within the Kondo hypothesis, the order of magnitude of the Kondo temperature $T_K$ can be obtained from a Curie-Weiss fit of the susceptibility: $\chi =C/(T+\theta)$, over a restricted temperature range. For sample $B-b$, one finds $\theta=$ 2.5 K when the analysis is performed from 10 K to 150 K, thus a $T_{K}$ ($\sim \theta$) lying in the Kelvin range. 
At this stage, the question is the following: How does the magnetic behavior 
evolve when one changes progressively the moment concentration from that 
of sample $B-b$, with presumably dominating Kondo couplings, to that 
of sample $R$ with pure RKKY couplings ? To answer this question, we performed a study for 22 samples with a $x$ varying by 2 
orders of magnitude spanning its range from that of sample $B-b$ to 
sample $R$. 

Up to now, we have put aside the specificity of the $Mn$ magnetism in the QC phases. But we must consider, even briefly, the very intriguing problem of the formation of effective moments in the present material. First, we recall that the $Al-Pd-Mn$ QC phase is thermodynamically stable only for a narrow range of compositions, containing a rather large amount of  $Mn$, about 8 at\%. But only a very few $Mn$ atoms appear to carry an effective spin. Interestingly enough for our study, the fraction of magnetic $Mn$ has been found to vary 
by orders of magnitude just from very slight composition changes
\cite{art_Hippert-2003}. This is even more interesting in our case since $Mn$ atoms are a component of the structure. Thus those $Mn$ atoms carrying an effective spin are not considered as 
a different chemical species. So the electronic structure 
could be very little affected by large variations of the moment 
concentration, which is a 
singular advantage that standard alloys cannot present. To give more details, let us recall that the Curie 
constant of $Al-Pd-Mn$ QC is found to be two 
or three orders of magnitude smaller than that calculated when all the Mn atoms 
are assumed to carry a 
spin $S=5/2$. For instance, the Curie constant of sample $R$ ($B-b$) is
17 (2000 respectively) times smaller than that expected if 
all the $Mn$ atoms carry a spin 5/2. The latter is calculated from the standard formula for the Curie constant of a paramagnet: $C=\alpha  xS(S+1)$, with $\alpha =N  
g^{2}\mu_{B}^{2}/3k_{B}$, $g=2$ ($N$ denotes the total number 
of atoms). Note that the only measurement of the Curie constant, $\propto xS(S+1)$, 
cannot provide separate values of $x$ and $S$. So, because the $x$ value is \textit{a priori} unknown in any sample, several studies must be carried to deduce the 
magnitude of $S$: One could get a $S$ value equal to about $5/2$ from the analysis of the field 
curvatures of the magnetization\cite{art_lasjaunias,Trans-SG}, taking into account the magnetic interactions. Consequently, the $x$ value must be small to explain the magnitude of $C$, which has been confirmed by NMR studies\cite{art_nmr2,art_nmr3} which locally probes the 
magnetism:  Most of the Mn 
atoms do not carry an effective electronic spin (at least at the 
time scale of the nuclear relaxation). In the present paper, we are interested in the interactions which affect the few $Mn$ atoms which do carry an effective spin. 

The paper is organized as follows. 
We present briefly our samples in 
Sec. \ref{Samples} and the linear susceptibility data in Sec. \ref{M,X}. In Sec. \ref{Kondo
effect}, we analyze the $\chi (T)$ with Kondo models (first with a single $T_{K}$ and subsequently with a 
broad $T_{K}$ distribution), which we extend to the magnetization in larger fields. The summary, discussion and possible microscopical pictures are developed in Sec. \ref{dis-2TK}  before the conclusion (Sec. \ref{conclusion}).

\section{Samples and experimental}
\label{Samples}

Let us recall that the quasicrystals are metallic phases which exhibit a long range order 
with a five-fold symmetry. In the case of the thermodynamically stable $Al-Pd-Mn$ QC phase, 
one can grow large sized single grains of high structural quality. 
It is very favorable for our study because the moment concentration is very sensitive 
to both the chemical composition and local atomic order \cite{art_Hippert-2003}. 
So the best is to avoid grain boundaries where 
 the moment concentration could be different from that in the bulk. We studied 22 samples issued from 15 pieces cut in 11 single grains (denoted by capitals) of slightly different 
composition (within 21.5-22.5 at\% of Pd and 
7.5-8.7 at\% of Mn). The grains were grown with the Czochralski method. We drew advantage of the sensitivity of the moment concentration 
to the annealing procedure to get successively "samples" of 
different moment concentration from the same QC piece, as described in Ref. 
\onlinecite{art_Hippert-2003}. For samples cut in grains $A$, $B$, 
$C$, $E$, $G$, $H$, one can refer to a previous paper
\cite{art_Hippert-2003} where many details are given about the preparation, 
the composition, the required precautions 
for cutting pieces of homogeneous composition, the heat treatments and the atomic structure. Here the sample notations are those used in 
Refs.\onlinecite{art_Hippert-2003} and \onlinecite {Sig-pre}. For 18 samples, partial 
magnetic studies have been previously published\cite
{art_Hippert-2003,art_lasjaunias,Sig-pre}, but with no systematic Kondo
analysis. The only sample which is not issued from a single grain is 
that denoted $R$. It is melt-spun  because of its composition, 
$Al_{71}Pd_{18}Mn_{11}$, out of the stability range. 
It is more magnetic than the present other 
samples and thus one expects the presence 
of grain boundaries to be of little importance, because 
the average moment concentration is large in this sample. 
We measured the magnetization and the 
susceptibility of all samples using SQUID magnetometers described in 
Ref.\onlinecite{art_Hippert-2003}. All the data presented in the 
following are corrected from the 
diamagnetic contributions in the way indicated in 
Ref.\onlinecite{art_Hippert-2003}.

\section{Magnetization and susceptibility}
\label{M,X}

\begin{figure}
\includegraphics[width=8cm]{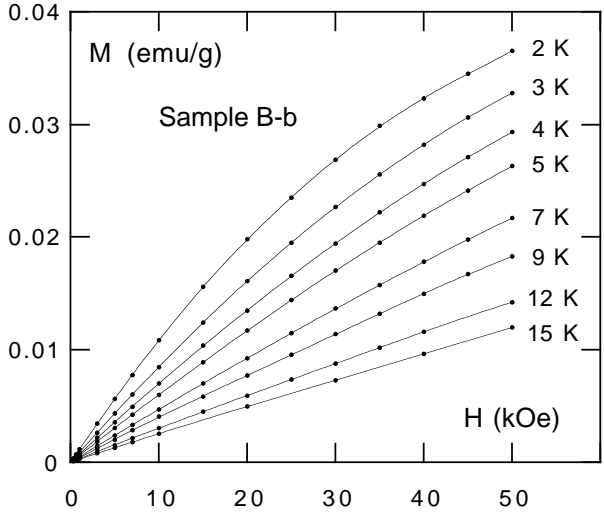}
\caption{Field dependence of the magnetization $M$ [corrected from the 
diamagnetic contribution (in emu/g): $\approx -4 \times 10^{-7}\times H$] for sample $B-b$ at
different constant temperatures ($T\geq 2$ K). The curves are guides for the
eye.} \label{fig0}
\end{figure}

\begin{figure}
\centering{\includegraphics[width=8cm]{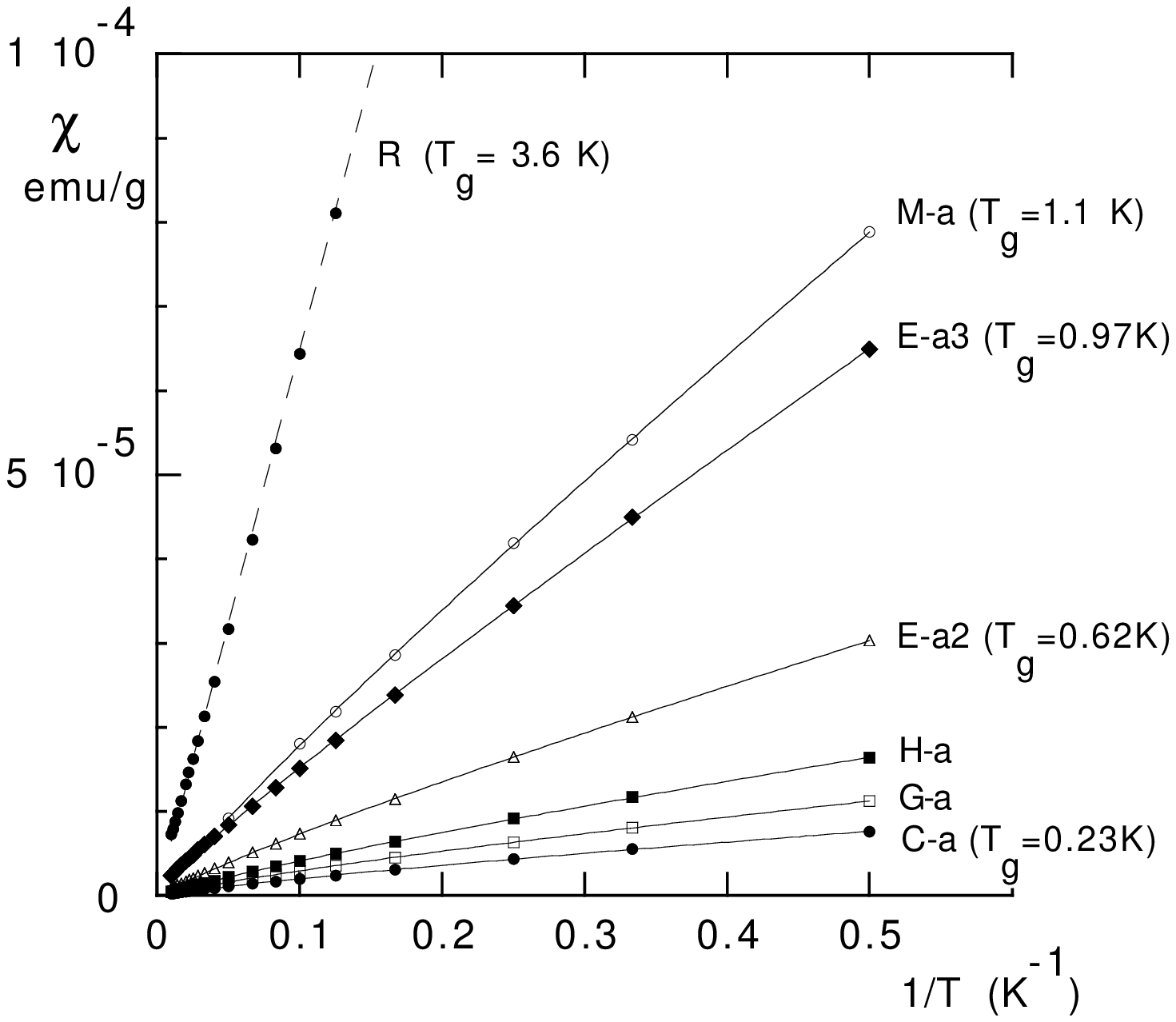}}
\vspace*{0.5cm}
\centering{\includegraphics[width=8cm]{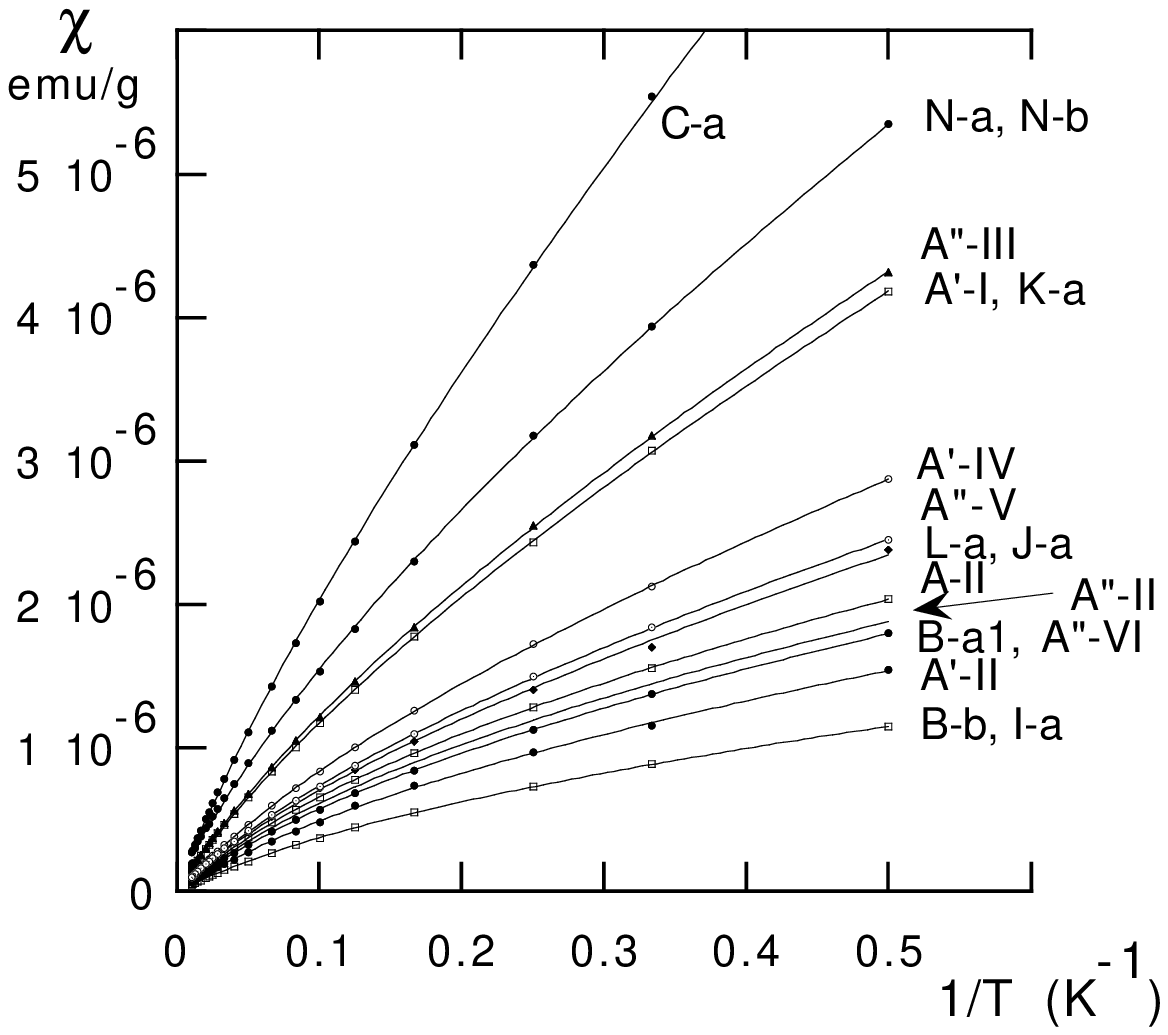}}
\caption{The d.c. susceptibility (for $T\geq 2K$) is represented in a diagram 
$\chi$ vs. $1/T$ for 22 samples. Here, $\chi=M/H_m$ (see text) with $H_{m}=$ 
1 kOe for all samples except for the more magnetic ones: 200 Oe for $E-a_{2}$ and 
$E-a_{3}$ and about 10 Oe for $M-a$ and $R$. Note the
different vertical scales required by the drastic sample-dependence of
the susceptibility. The solid curves are Kondo fits within the 
one-$T_{K}$ model}
\label{fig2}
\end{figure}

\begin{figure}
\centering{\includegraphics[width=8cm]{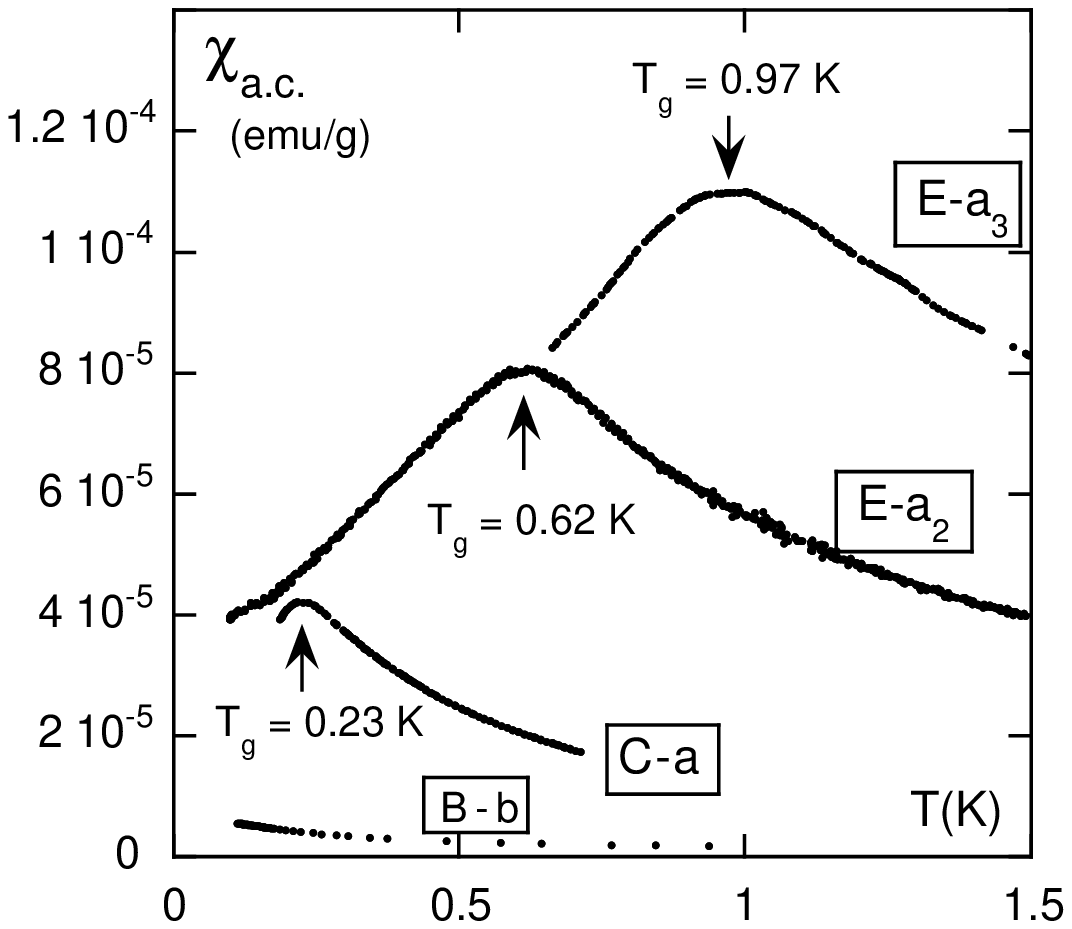}}
\vspace*{0.5cm}
\caption{Temperature dependence of the a.c. susceptibility $\chi_{a.c.}$ for
four samples ($h_{a.c.}$ = 1 Oe, frequency 1 Hz). Note that $T_{g}$ values of
1.1 K and 3.6 K have been reported for samples $M-a$ and $R$ in
Ref.\onlinecite{art_lasjaunias} }
\label{fig5}
\end{figure}

The main features of the magnetization 
$M$ of sample $B-b$ are shown in Fig.\ref{fig0} where we represent its field dependence 
measured up to 50 kOe ($\mu_0H\,=$ 5 Tesla) at different constant temperatures $T\geq 
2K$. The $M(H)$ curves follow the behavior of a magnetization due to localized moments in a 
paramagnetic regime. Indeed, first, the initial slope, that is the linear susceptibility $\chi$, decreases continuously at increasing temperatures. Second, the magnetization is $H$-linear up to fields larger and larger 
at increasing temperatures. It allows one to escape systematic measurements of $M(H)$ at fixed temperatures to deduce $\chi (T)$. Indeed, it is sufficient to measure the magnetization $M(H_{m},T)$ as a function of the temperature in a constant measurement field $H_{m}$. The condition to identify $ \chi (T)$ for $T\geq 2$ K with $M(H_{m},T)/H_{m}$ is simply that $H_{m}$ lies in the initial field range where $M(H)$ is 
field-linear to better than 1\% at 2 K. We used this method to obtain the $\chi$ data of all the samples shown in Fig.\ref{fig2}. 
For some selected samples, we have also measured the magnetization in 
d.c. fields and the low frequency (1 Hz) a.c. susceptibility $\chi_{a.c.}$ down to 100 mK. For the very weakly magnetic sample $B-b$, we could not detect any a.c. susceptibility peak down to 110 mK. 
For more magnetic samples, namely $C-a$, $E-a_{2}$ and $E-a_{3}$, the $\chi_{a.c.}(T)$ is observed to exhibit a 
cusp at a sample dependent temperature denoted $T_{g}$: See Fig.\ref{fig5}. 
These data add 
to those previously published\cite{art_lasjaunias} for more magnetic samples: $M-a$ ($T_{g}=1.1$ K) and $R$ ($T_{g}=3.6$ K). 
We also studied the 
zero-field cooled and field cooled magnetizations obtained in 
d.c. fields below $T_{g}$ for samples $E-a_{2}$ and 
$E-a_{3}$: We recovered the same magnetic hysteresis features (not 
shown here) as those observed for sample $R$ presented in Fig.\ref{fig00}. 

\begin{figure}
\centering{\includegraphics[width=8cm]{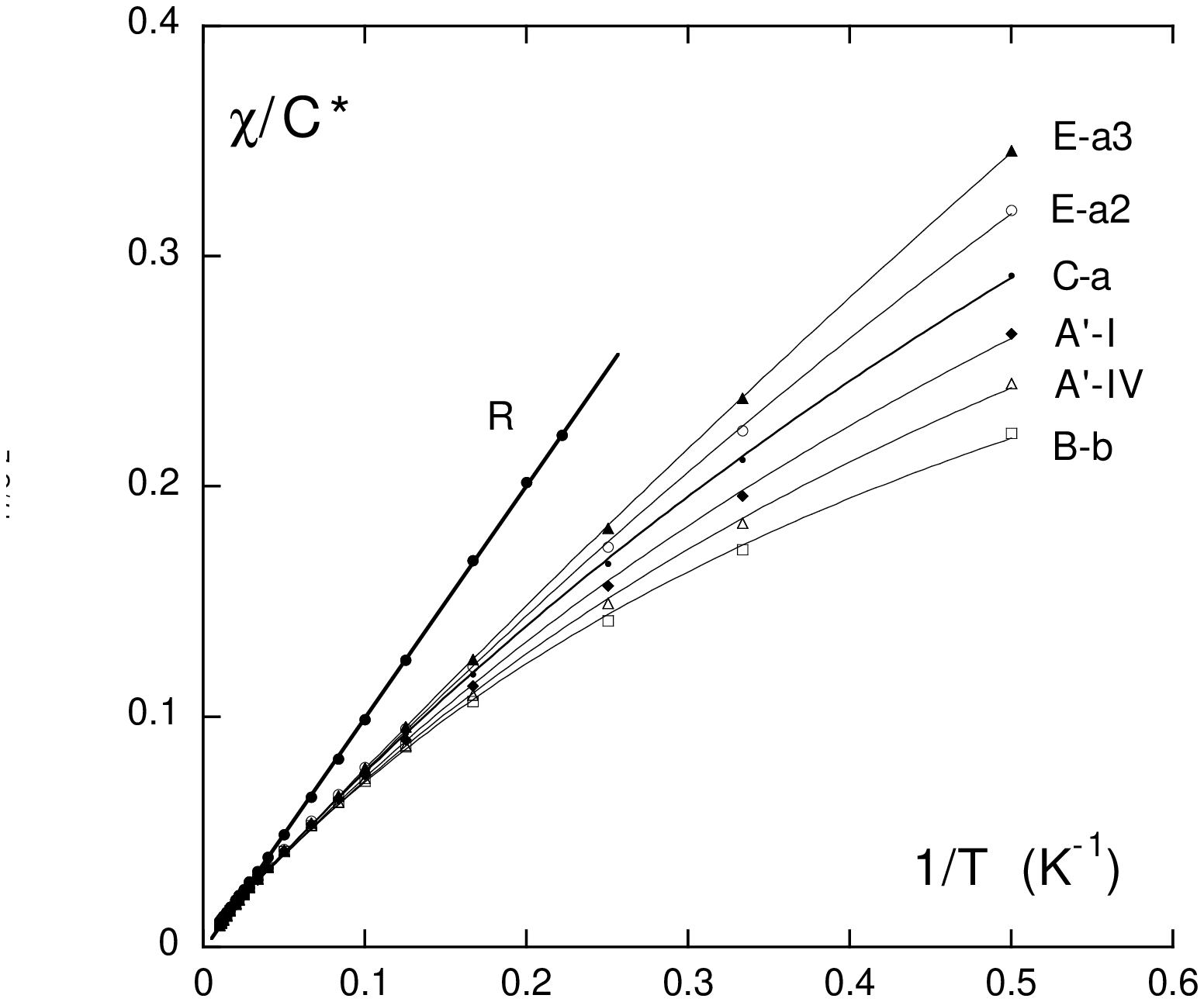}}
\caption{Normalized susceptibility $\chi  /C*$ vs. 1/T. Only a few curves are 
shown for sake of clarity.}
\label{fig03}
\end{figure}

Let us summarize our results. First, we have studied samples which allow us to  
follow the evolution of the magnetic behavior when changing the 
moment concentration over two orders of magnitude: From that of $B-b$ to that of 
sample $R$. Second, the same sample exhibits continuous $\chi (1/T)$ curvatures at moderate and large temperatures, here attributed to Kondo effects, and a spin glass transition at low temperature, that we showed for several samples. Third, we observe an opposite variation of the relative weights of the Kondo and RKKY effects. On one hand, the $T_{g}$ is found to increase for
samples which are more and more magnetic. Thus with increasing moment concentration: From 
0.23 K for sample $C-a$ up to 3.6 K for sample $R$ (See Fig.\ref{fig5} in connection with Fig.\ref{fig2}). This feature 
agrees well with the behavior expected for RKKY spin-glasses. On the other hand, the $\chi(1/T)$ curvatures are found to decrease with increasing moment concentration. This can be observed directly from the evolution shown in Fig.
\ref{fig2} and appears even clearer in Fig.\ref{fig03} where we plotted $\chi /C^{*}$ vs. $1/T$. Here, $C^{*}$ is a parameter calculated for each sample in order to get  the best superposition as possible of the 
$\chi(1/T)$ curves in the [20K-100K] $T$-range. Although purely phenomenological, this plot shows that the curvature of $\chi(1/T)$ is more and more pronounced when the 
sample is less and less magnetic. Thus when the single-moment limit is 
approached: Compare Fig.\ref{fig03} and \ref{fig2}. Surely, this observation reinforces our hypothesis of Kondo effects being at the origin of the $\chi(1/T)$ curvatures. But in turn, this shows that the RKKY 
couplings end by killing the $\chi(1/T)$ curvatures when the moment 
concentration is increased. 
At this stage, one needs a quantitative data analysis in order to understand the 
relative role of the RKKY and Kondo interactions. This is the aim of the next section.

\section{Quantitative analysis}
\label{Kondo effect}

We attributed the $\chi(1/T)$ curvatures observed above 2 K, thus above $T_{g}$, to Kondo effects. So, we chose to analyze these curvatures in a very simple way, i.e. fitting our data with a pure Kondo model. Only in a second step, we shall discuss the evolution of the values of the fitting parameters with the moment concentration and, thus, in view of the RKKY coupling effects. In a  first attempt (Sec. \ref{T>2K}), we use the simplest Kondo model with a single interaction energy 
scale (denoted $T_{Keff}$) for a given sample. Subsequently (\ref{two-TK effect}), we extend the analysis when assuming the existence of a broad  $T_{K}$ distribution.

\subsection{One-$T_{K}$ Kondo analysis}
\label{T>2K}

We first analyze the 
$\chi(T)$ data using a 
genuine Kondo model implying only one energy scale, $T_K$. Instead of fitting these data with a 
Curie-Weiss law, we use the results\cite{Noz-Blandin,Desgranges,Sacramento} 
obtained from the $n$-channel Kondo model, assumed to be suited to the case $S>1/2$. In case of $n=2S$, the Kondo susceptibility $\chi_K(T)$ for $S>1/2$ behaves as that for $S=1/2$. In particular, it reaches a saturated value $\chi_K(T=0)$ ($\propto 1/T_{K}$) when $T\rightarrow 0$, in strong contrast with the Curie 
susceptibility which diverges as $1/T$. With the definition of $T_K$ given in Appendix \ref{XK}, one obtains Eq.\ref{eq6} and thus: 
 \begin{equation}
\ \chi_{K}(T,T_K)=\frac{0.616\times \alpha xS}{T_{K}}  F(\frac{T}{T_{K}}), {\rm with}\, F(0)=1
\label{eq1}
\end{equation}
where $\alpha =N g^{2}\mu_{B}^{2}/3k_{B}$ and where we put $S=5/2$. $F(T/T_K)$ has been found\cite{Desgranges,Sacramento} to be almost independent on $S$ at low and moderate $T/T_K$. Thus it can be identified with that one obtained for $S=1/2$. In the present section, we neglect the deviations (commented in Appendix \ref{XK}) of $F(T/T_K)$ from the $S=1/2$ results at larger $T/T_K$.  So, we fit the $\chi(T)$ data at all temperatures, using the $F(T,T_K)$ (shown in Fig.\ref{fig4}) calculated numerically from the $S=1/2$ results\cite{Wilson-1,Wilson-2} and as explained in Appendix \ref{XK}. The fit is quite satisfactory for each sample over two temperature decades, from 2 K up to 100 K. In 
Fig.\ref{fig2} the solid curves are the fitting
curves. For each sample, 
the fit provides the values of two parameters: 
The Kondo temperature 
which governs the $\chi(1/T)$ 
curvature and the moment concentration which governs the susceptibility amplitude. The present analysis being a first attempt, we 
cannot assert however at this stage that the latter parameters represent really the actual Kondo 
temperature and the exact moment concentration. 
Thus, in the following, we call them an \textit{effective} Kondo 
temperature $T_{Keff}$ and an \textit{effective} moment concentration $x_{eff}$ respectively. Once the $T_{Keff}$ and 
$x_{eff}$ are deduced from the susceptibility of each sample, we can replot the $\chi (T)$ data of all the samples in the scaling diagram: $ \chi \times 
T_{Keff}/(0.616\alpha Sx_{eff}) $ $vs.$ ${\rm Log} (t)$ with $t=T/T_{Keff}$, which proceeds from formula Eq.\ref{eq1}:  See Fig.\ref{fig4}. Then, all the data are superposed on the theoretical $F(t)$ curve.
Next, we comment on the sample dependence of $T_{Keff}$ and 
$x_{eff}$.
We found a $T_{Keff}$ varying widely, by 
a factor 10, when the $x_{eff}$ value varies by a factor 40, from $1.4\times 10^{-4}$ for sample $B-b$ 
to $5.4\times 10^{-3}$ for
sample $M-a$: See Fig.\ref{fig3}. Strikingly enough, 
the $T_{Keff}$ seems to follow a power law over the present $x_{eff}$-range: 
$T_{Keff}\sim x_{eff}^{-0.55\pm 0.05}$. 
Qualitatively, the 
$T_{Keff}$ variation is not really a surprise: The $T_{Keff}$ value 
reflects the amplitude of the $\chi (1/T)$ curvatures, which was found to decrease at increasing moment concentration. Here, the new result is that we have quantitative $T_{Keff}$ data. 
\begin{figure}
\centering{\includegraphics[width=8cm]{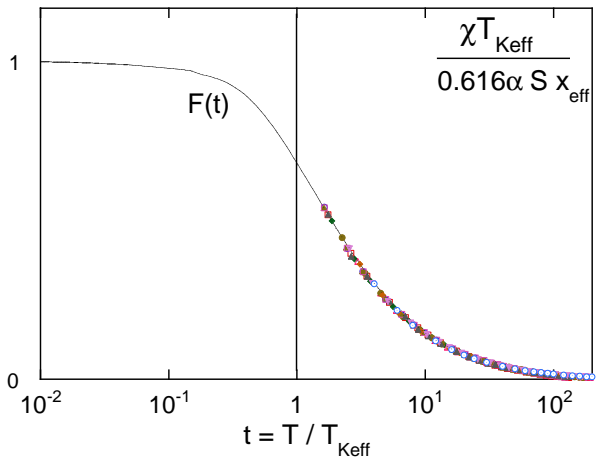}}
\caption{Kondo scaling for all the samples (except $R$) for
$T\geq2K$} 
\label{fig4}
\end{figure}
\begin{figure}
\centering{\includegraphics[width=8cm]{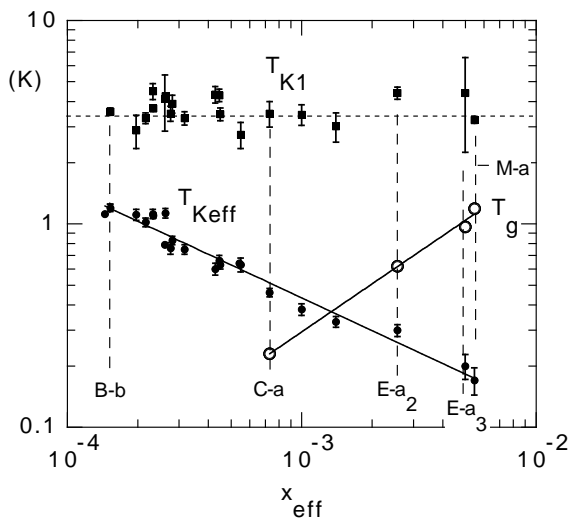}}
\caption{Log-log plot of $T_{Keff}$ (solid circles) and $T_{K1}$ (solid squares) vs. the parameter
$x_{eff}$. The spin-glass 
temperature $T_{g}$ (open circles) is reported for four samples. The 
solid lines are power fits: $T_{Keff}\propto x_{eff}^{-0.55}$, $T_{g}\propto x_{eff}^{0.8}$.} 
\label{fig3}
\end{figure}
In conclusion, it seems possible to fit successfully the $\chi (T)$ data 
above 2 K with a pure Kondo law. But one has to investigate the possible origins of the strong dependence of $T_{Keff}$ on $x_{eff}$. For that, let us assume in the following discussion that $x_{eff}$ reflects the real moment concentration $x$. Let us first stick to our starting hypothesis of a single  energy scale, the $T_{Keff}$, governing the $\chi (T)$ behavior outside the spin-glass ordered phase. Then, one can imagine two different attractive scenarios, each accounting for the $T_{Keff} (x_{Keff})$ variation. In the first one, $T_{Keff}$ is a standard Kondo temperature but renormalized by the RKKY couplings. This looks reasonable since the 
value of $T_{g}$ can be comparable with that of $T_{Keff}$: See Fig.\ref{fig3}. In addition, one observes  the $T_{g}$ to vary with $x_{eff}$ ($T_{g}\propto x_{eff}^{0.8\pm 0.1}$) in a way opposite to that of $T_{Keff}$ ($\propto x_{eff}^{-0.55}$). So, one can conjecture $T_{Keff}$ to be a function of both the $bare$ (no RKKY) Kondo temperature 
$T_{K0}$ and the RKKY  (no Kondo) energy scale $T_J$ ($\propto x_{eff}$) in such a way that $T_{Keff}$ decreases from $T_{K0}$ at vanishing $x_{eff}$ to zero when $x_{eff}$ increases sufficiently. 
The second scenario is that of the so-called exhaustion principle\cite{Noziere-2}. Here, the RKKY couplings above 
$T_{g}$ are ignored, but one is interested in the number $n_e$ of conduction electrons available to screen the Mn spins. These electrons  are of energy lying within $T_{K}$ from the Fermi level:  $n_e$ $\approx \rho \times T_{K}$. To treat our case, we use the value of the DOS $\rho $ deduced from specific heat results
\cite{art_lasjaunias,art_specif1,Swenson} and put a $T_{K}$ equal to 1.2 K which is our largest $T_{Keff}$ value. Then we find a $n_e$ which is, depending on $x_{eff}$, 70 to 2500 times smaller than the number $2S\times Nx_{eff}$ which should be necessary to screen all the Mn spins. In that case, Nozi\`eres\cite{Noziere-2} proposed to introduce a coherence temperature $T_{coh}$ below which the standard low-$T$ Kondo regime occurs. This $T_{coh}$, which could be the $T_{Keff}$ in our case, is predicted to equal at most $T_{K}\times n_e /2SNx_{eff}$, where 
$T_{K}$ is the actual Kondo temperature (no exhaustion). It is a prediction ($T_{coh}\propto 1/x_{eff}$) which is qualitatively in agreement with our result:  $T_{Keff}\propto x_{eff}^{-0.55}$.

\begin{figure}
\centering{\includegraphics[width=8cm]{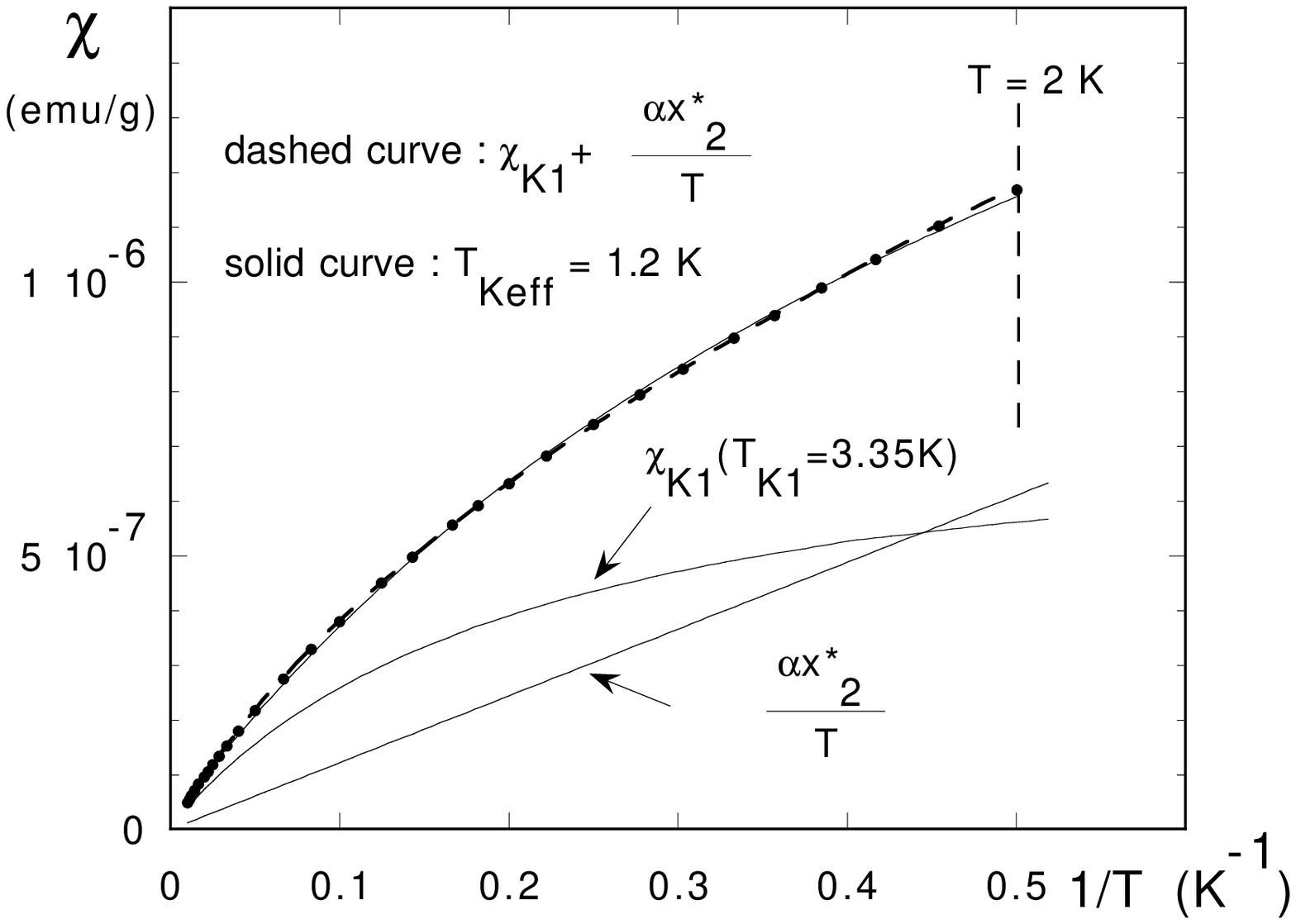}}
\vspace*{0.5cm}
\centering{\includegraphics[width=8cm]{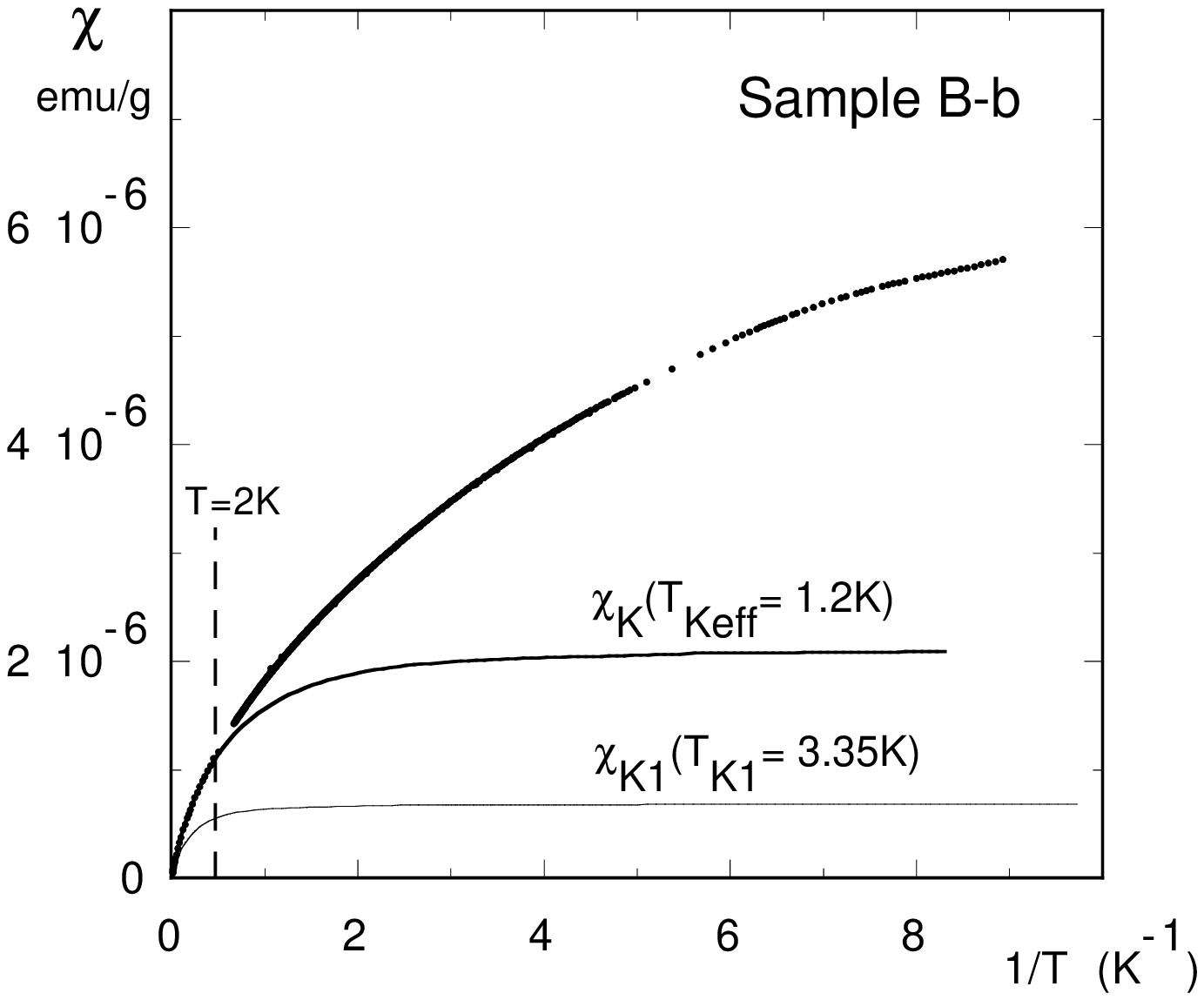}} 
\caption{Susceptibility of sample $B-b$ vs. $1/T$: down to 2 K in the upper diagram and down to 0.1 K in the lower diagram. The susceptibility below 2 K was obtained from d.c. (d.c. field 100 Oe) and a.c. ($h_{ac}=1$ Oe) measurements. Here we show the Kondo curve fit obtained above 2 K when using a single-$T_{K}$ (solid curve) and  a two-$T_{K}$ model (dashed curve).  Also, we show the $\chi_{K}(T,T_{Keff})$ fitting curve extrapolated down to 0.1 K, the $\chi_{K1}(T)$ (both diagrams) and $\chi_{K2}(T)=\alpha Sx_2*/T$ (upper diagram) curves calculated with the $x_{1}$, $x_2^{*}$ and $T_{K1}$ values for this sample.}
\label{fig2d}
\end{figure}

\begin{figure}
\centering{\includegraphics[width=8cm]{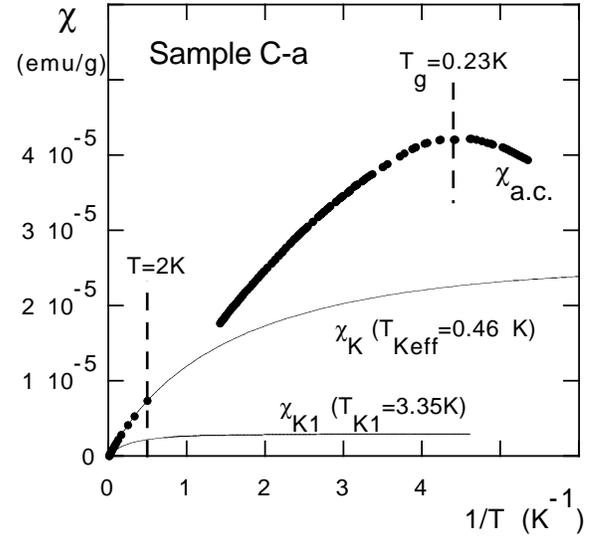}}
\caption{Same as Fig.\ref{fig2d} but for sample $C-a$. The data below 2 K are a.c. susceptibility ($h_{ac}=1$ Oe).} 
\label{fig7}
\end{figure}

The common feature of the two previous scenarios is a susceptibility reaching its saturation value below $T_{Keff}$. However, up to now, we only explored the temperature regime above 2 K while we found $T_{Keff}$ values always smaller than 2 K (See Fig.\ref{fig3}). So we missed the low-$T$ regime i.e. $T/T_{Keff}<1$ (See Fig.\ref{fig4}) and thus an important data set. However, to extend the Kondo study below $T_{Keff}$ requires a $T_g$ much lower than $T_{Keff}$. See Fig.\ref{fig3}: Sample $B-b$ is a good candidate for such a study. Its $T_{Keff}$ is large (1.2 K) and its $T_{g}$, if it exists, is well below 0.1 K (Sec. \ref{M,X}). In the diagram $\chi$ vs. $1/T$ in the upper part of Fig.\ref{fig2d}, we show the data obtained down to 2 K together with the Kondo fit which provides the values of $x_{eff}$ and $T_{Keff}$ for this sample. In the lower part of Fig.\ref{fig2d} we show the same data plus those obtained down to 110 mK in the same type of diagram but now extended up to $1/T=10\ K^{-1}$. Also, we show the previous fit $\chi_{K}(T,T_{Keff})$ that we extrapolated down to 0.1 K. One could expect the $\chi (T)$ to reach its saturation value below 
$T_{Keff}$ as suggested from the extrapolated fitting curve. But it is obviously far from being the case: The low-$T$ data are well above this curve. The same feature is observed for a more magnetic sample, $C-a$, of smaller $T_{Keff}$ ($0.46\ K$) but nevertheless  larger than $T_g$ ($0.23\ K$): See Fig.\ref{fig7}. Such results question the validity of our starting hypothesis of a simple ($n=2S$)-Kondo susceptibility. Two scenarios within the $n$-channel Kondo model can explain the absence of saturation of $\chi (T)$ when $T$ goes to 0: That one where $n\neq 2S$ with a single $T_K$ and that one where $n=2S$ but with a broad $T_K$ distribution. For the $n\neq 2S$ scheme, two cases are considered\cite{Sacr-1991}. In the undercompensated case ($n<2S$), the spin is only partially compensated: It remains a spin $S'=S-n/2$ at low temperature, obeying a Curie behavior ($\chi \propto 1/T$). In the overcompensated case ($n>2S$), the $\chi (T)$ is expected to vary as $(T/T_K)^{\tau - 1}$ at low temperature with $\tau=4/(n+2)$. In the case of sample $B-b$, we find a $\chi (T)$ varying as $T^{u}$ with $u$ decreasing slightly with the temperature, equal to nearly $0.55$ for 0.14 K $<T<$ 0.5 K, from which one deduces $n \approx 5$ (hence a spin which has to be assumed smaller than 5/2). However, these ($n\neq 2S$)-models suppose the susceptibility to be governed by a $x$-independent $T_K$ and thus to be simply proportional to the moment concentration, which is inconsistent with our results. Let us consider the alternative hypothesis: $n=2S$ but with a broad $T_K$-distribution $P(T_{K})$. Then, the $\chi
(1/T)$ curvatures observed well above 2 K could be accounted for by the
existence of large Kondo temperatures whereas the absence of saturation of $\chi
(T)$ at very low temperatures could be explained by the existence of very low
$T_{K}$ values. In the present case, one has to assume the shape of $P(T_K)$ to depend on the 
moment concentration to account for our results. Indeed, in this case, the evolution of $P(T_{K})$ with the moment concentration leads to that of our fitting parameter $T_{Keff}$ because the latter is  obtained from fits carried over a fixed temperature window (2 K-100 K). This working hypothesis is examined in the next section. 

\subsection{Kondo analysis with a $T_{K}$ distribution}
\label{two-TK effect}

Let us summarize the main result of Appendix \ref{P(Tk)}. A Kondo 
susceptibility $\chi_K (T, T_K)$ averaged over a broad $T_{K}$ distribution can be 
described by the sum of a limited number of single-$T_{K}$ Kondo susceptibilities $\chi_{Ki}(T,T_{Ki})$. 
For instance, using formula Eq.\ref{eq1} for $\chi_K (T,T_K)$, one can write: $\chi _{Ki}(T,T_{Ki})= x_iS\times (0.616 \alpha/T_{Ki}) \times F(T/T_{Ki})$ where the $x_i$ are parameters having the dimension of a moment concentration. The number of terms in the sum depends on the width of the $T$-range used for the 
analysis. If the $T$-range spreads over two $T$-decades, one term is sufficient, implying a Kondo temperature that we identify with our previous $T_{Keff}$. If the $T$-range is extended to a third decade, two terms are necessary to describe the susceptibility: 

\begin{equation}
\ \chi(T) =\chi _{K1} (T, T_{K1}) + \chi _{K2} (T, T_{K2})\hspace{0.25cm} {\rm with} \hspace{0.25cm} T_{K1}\gg T_{K2}
\label{eq2}
\end{equation}
As shown in Appendix \ref{P(Tk)}, the value of $T_{Keff}$ lies 
in between those of $T_{K1}$ and $T_{K2}$. 
To fit susceptibility data with the formula Eq.\ref{eq2} is difficult because one deals with $F(T/T_{K1})$ and $F(T/T_{K2})$ which are not simple analytic functions. However, in view of the properties of $F(t)$, we can simplify the analysis in two cases: $T<T_{K1}$ and $T\gg T_{K2}$. Indeed, for $T<T_{K1}$, the 
$\chi_{K1}$ should have reached its saturated value and thus can be replaced by a constant in formula Eq.\ref{eq2}. For $T\gg T_{K2}$, the $\chi_{K2}$ lies in the high temperature Kondo regime, where the Kondo susceptibility is given by a Curie term with logarithmic corrections. Consequently, one can write: $\chi_{K2}= \alpha  x_{2}S q/T$ when the explored temperature range is restricted. The value of the parameter $q$ depends on how the high temperature Kondo susceptibility is parametrized (See Appendix \ref{XK}). In summary, writing $\phi (T,T_{K})=\alpha  (0.616/T_K) F(T/T_K)$ we propose the following formula:

\begin{subequations}
\begin{gather}
T<T_{K1}:\ \chi(T) =Constant + x_{2}S  \phi (T,T_{K2})\label{eq3a} \\
T\gg T_{K2}:\ \chi (T) \approx x_{1}S  \phi (T,T_{K1}) 
+ \alpha  qx_{2}S/T ,\label{eq3b}
\end{gather}
\end{subequations}
Hereafter, $qx_{2}$ will be denoted by $x_{2}^{*}$. 
Let us apply this analysis for sample $B-b$. First, the value of $T_{K1}$ should be larger than 1.2 K, the previous estimate of $T_{Keff}$. Then, we can fit the data obtained below 1 K ($<T_{K1}$) with formula Eq.\ref{eq3a}. The fit provides a value for $T_{K2}$ equal to about 100 mK. Turning now to the analysis of the data obtained at $T\geq 2K$, we note that the $T/T_{K2}$ values are huge in that $T$-range: From 20 up to $10^{3}$ for $T$ ranging from 2 K to 100 K. It follows that we can apply the formula Eq.\ref{eq3b} to fit the data obtained above 2 K. The obtained fit is even better than that carried out with a  single $T_{K}$ (See Fig.\ref{fig2d}), due to the addition of the third fitting parameter $qx_{2}$. The fit provides a value of $T_{K1}$ equal to 3.35 K. Thus, we have explored a $T/T_{K1}$ range, from 0.6 (for $T=2$ K) to 30 ($T=100$ K), which allows us to neglect the deviations of $F(T/T_{K1})$ in $\phi (T,T_{K1})$ from the $S=1/2$ results (See Appendix \ref{XK}). For the other samples ($C-a$, $E-a_2$, $E-a_3$) studied below 2 K, we also observed strong deviations of their susceptibility below 2 K from the extrapolated curve fit $\chi_K (T,T_{Keff})$, as we showed for sample $C-a$ in Fig.\ref{fig7}. This stimulates us to reanalyze all the data, even restricted to the [2 K-100 K] $T$-range, with a two-$T_{K}$ model and to compare the values of the fitting parameters with those ($T_{Keff},\ x_{Keff}$) deduced previously. It is the subject of the following section. 

\subsubsection{Susceptibility}
\label{2-TK-X}

\begin{figure}
\centering{\includegraphics[width=8cm]{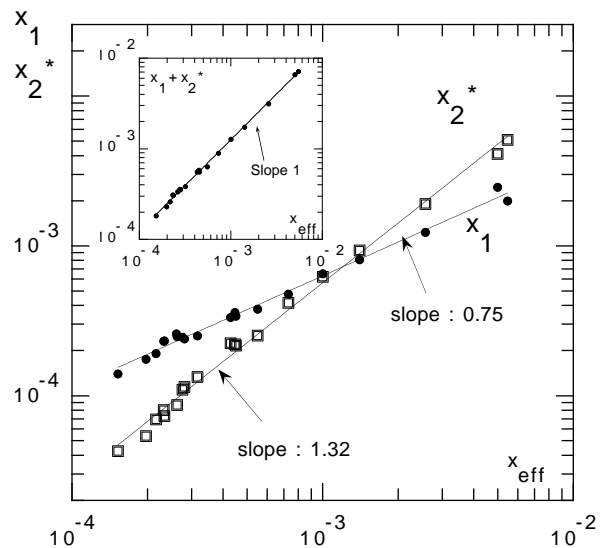}}
\caption{$x_{1}$ and $x_{2}^{*}$ vs. $x_{eff}$ and $x_{1}+x_{2}^{*}$ vs. 
$x_{eff}$ (inset) in log-log diagrams}
\label{Ai-xeff}
\end{figure}

\begin{figure}
\centering{\includegraphics[width=8cm]{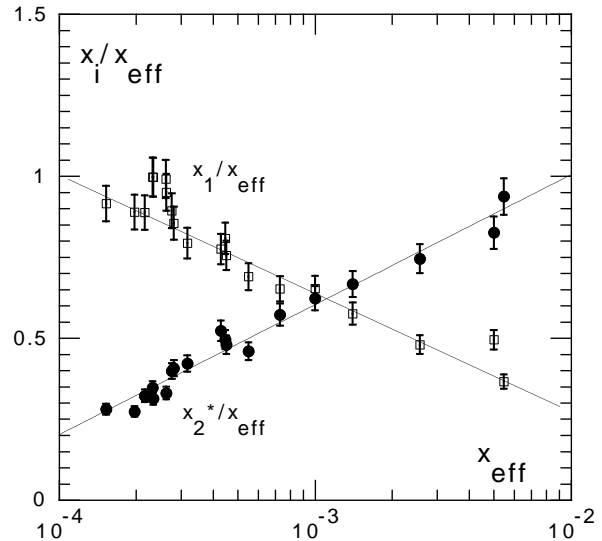}}
\caption{$x_{1}/x_{eff}$ and $x_{2}^{*}/x_{eff}$ vs. $x_{eff}$ in a semilog diagram. The 
error bars follow from the relative errors for the determination of 
$x_{1}$, $x_{2}^{*}$ and $x_{eff}$. The 
lines are guides for the eye}
\label{Ai-norm-xeff}
\end{figure}

For all samples, the one-$T_K$ Kondo fit of the susceptibility over two $T$-decades was good. Consequently, an even better fit of the same data using a two-$T_K$ model (formula Eq.\ref{eq3b}) is not surprising. Actually, the surprise comes from the remarkable and very simple result provided by the latter fit: The $T_{K1}$ value is found to be constant within our accuracy, i.e. independent on the moment concentration in contrast with the previous $T_{Keff}$ (see Fig.\ref{fig3}). More spectacular is that it does so up to the limit where the $\chi (1/T)$ curvatures have almost disappeared (case of sample $M-a$ shown in Fig.\ref{fig2}). We propose a value of 3.35 $\pm$ 0.15 K for $T_{K1}$, which is that obtained from data on our largest size samples. Indeed, the accuracy on $T_{K1}$ is mainly determined by that of the susceptibility data which is
the better the larger sample mass is (hence the error bars in Fig.\ref{fig3}). Another 
result concerns the amplitude parameters $x_{1}$ and $x_{2}^{*}$ deduced from Eq.\ref{eq3b} assuming $S=5/2$. First, the 
sum of $x_{1}$ and $x_{2}^{*}$ is found to be proportional to the 
$x_{eff}$ previously deduced: See the diagram in the inset of Fig.\ref{Ai-xeff}. More precisely, we find $x_{1}+x_{2}^{*}=1.32\times  x_{eff}$. Second, both $x_{1}$ and $x_{2}^{*}$ increase with $x_{eff}$, but at a different rate: See Fig.\ref{Ai-xeff}. About the $x_{1}+x_{2}^{*}\sim x_{eff}$, we find the result obtained in Appendix \ref{P(Tk)} from an over-simplified $T_K$-distribution: A two-$T_{K}$ analysis provides the same information than the one-$T_{K}$ analysis about the moment concentration. Let us take $x_{eff}$ as the moment concentration, at least in the following comment of the evolution of  $x_{1}$ and 
$x_{2}^{*}$ with $x_{eff}$. Quantitatively, both $x_{1}$ and 
$x_{2}^{*}$ are found to follow power laws over the 
present $x_{eff}$-range, but with a different exponent value: $x_{1}\propto x_{eff}^{0.75}$ and 
$x_{2}^{*}\propto x_{eff}^{1.3}$ as shown in Fig.\ref{Ai-xeff}. 
Then, we can deduce the evolution of the normalized distribution $P(T_{K})$ with the moment 
concentration $x_{eff}$: The weight of the susceptibility term associated to $T_{K1}$ is given by $x_1/x_{eff}$, while that, $x_{2}/x_{eff}$, associated to $T_{K2}$ is reflected by the $x_{2}^{*}/x_{eff}$ value. These weights are represented as functions of $x_{eff}$ in Fig.\ref{Ai-norm-xeff}.

\begin{figure}
\centerline{\includegraphics[width=8cm]{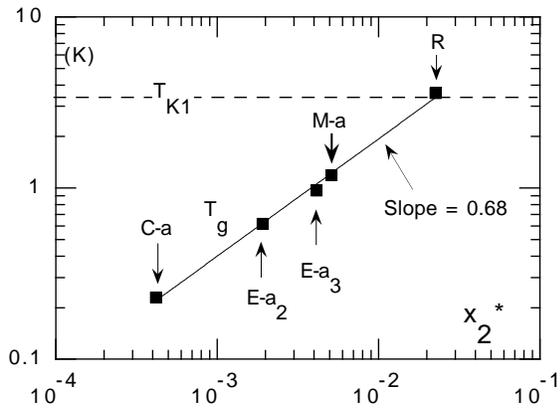}}
\vspace*{0.5cm}
\caption{$x_{2}^{*}$ dependence of $T_{g}$ in a log-log diagram. The $x_{2}^{*}$ of sample $R$ is deduced from  Eq.\ref{eq3b} where we put $x_1=0$ since no Kondo $\chi (1/T)$ curvatures can be detected for this sample. The solid line is a power curve fit. The dashed horizontal straight line indicates the value of $T_{K1}$} \label{figTg(x2)}
\end{figure}

At this stage, we can draw some interesting and simple conclusions. First, we could split the  susceptibility into two contributions, each exhibiting a specific behavior. One ($\chi_{K1}$) is Kondo-like with a Kondo temperature ($T_{K1}$) independent on the moment concentration. The other contribution ($\chi_{K2}$), to which we attribute an almost vanishing Kondo temperature, follows nearly a Curie law above 2 K. The more striking result is that the behavior of the two contributions holds with the same characteristics when the moment concentration is increased up to that one where the Kondo contribution disappears (sample $R$). Let us focus now on the (almost) non Kondo contribution to the susceptibility.
As shown below, it behaves as the susceptibility of a RKKY spin-glass containing a spin concentration proportional to $x_2^{*}$. First, $\chi_{K2}(T)\propto x_2^{*}/T$ above 2 K. Second, one finds $T_{g}$  obeying a power law of $ x_{2}^{*}$: $T_{g}\propto (x_{2}^{*})^{0.68}$ and, strikingly enough, it does so over almost two decades of $x_{2}^{*}$ when the data for sample $R$ are included: See Fig.\ref{figTg(x2)}. It  is because of this very large $x_{2}^{*}$ range that we can deduce the exponent value rather accurately ($0.68 \pm  0.02$) taking into account the relative precision on the ($T_g$, $x_{2}^{*}$) values. A similar  power law governs the $T_{g}$ dependence on the moment concentration in purely RKKY spin-glasses. Moreover the exponent value equals that found for the canonical $Cu-Mn$ spin-glass\cite{art_lasjaunias}. To complete the panorama of the susceptibility features, let us focus on the disappearance of the $\chi (1/T)$ curvatures. It occurs for sample $R$ while $\chi (1/T)$ curvatures are still present for sample $M-a$ which is only 3 times less magnetic (See Fig.\ref{fig2}). Thus sample $R$ can be taken as a crossover sample: Its moment concentration could be the minimum one necessary to observe RKKY couplings dominating fully. Another interesting feature is that the $T_{g}$ becomes comparable with $T_{K1}$ when the concentration is increased up to that of sample $R$: See Fig.\ref{figTg(x2)}. We shall discuss these two points in Sec. \ref{dis-2TK}. 
In summary, our analysis gives simple results in the zero-field limit (susceptibility analysis). The question is to know wether such an analysis holds when moderate or even large fields are applied. This is the object of the next section.

\subsubsection{Magnetization $M(H)$ at 2K}
\label{Mag-2K}

In a general way, one can write the magnetization as: $M=Nx\times g\mu_{B} S \times P(H,T,T_J, T_{K} \ldots)$. Here, $x$ and $P$ are the actual moment concentration and average spin polarization respectively. $P$ varies from 0 to 1 when the field is 
increased from 0 up to values large enough to overcome the temperature 
and interaction effects. Here, the largest characteristic 
temperature is $T_{K1}$. Then, because the maximum field attainable in 
our magnetometers equals 70 kOe, the available maximum value of $\mu 
_{B}H/k_{B}T_{K1}$ equals 1.4. 
Surely, this value is not large enough for a complete study of the 
magnetization up to its saturation value but it is expected to be 
sufficiently large for generating marked $M(H)$ curvatures suitable for 
a meaningful analysis. Formally, it is possible to decompose the magnetization of any sample at a given field and temperature as the sum of two terms:  
\begin{equation}
\ M(H,T)= Ng\mu_{B}S  [x_{1}P_{1}(H,T,T_1^{*})+x_{2}P_{2}(H,T,T_2^{*})]
\label{eqM1}
\end{equation}
where $x_1$ and $x_2$ have the values deduced from the susceptibility analysis. Here, we consider the more general formula for the $M(H,T)$ decomposition above $T_g$. We only assume the $T_i^{*}$ to be characteristic energy scales for the interactions in the presence of a field and we attribute no explicit physical meaning to the $P_i$. In the zero field limit where $M=\chi H$, formula Eq.\ref{eqM1} is equivalent to Eq.\ref{eq2} with $T_1^{*}=T_{K1}$ and $T_2^{*}=T_{K2}$. 
But we have no particular reason to assume that the same simple result holds in large fields, especially when $\mu_BH/k_B$ becomes comparable to $T_{K1}$. Surely, one can use formula Eq.\ref{eqM1} to fit the magnetization data for large fields but, then, each $T_i^{*}$ could depend on the variable parameters, i.e. $H$, $T$ and the moment concentration. To clarify this point, we present an analysis of the magnetization measured at 2 K. This temperature has the advantage to lie in-between $T_{g}$ and $T_{K1}$. Indeed this allows us to examine $P_{1}(H)$ in the more relevant, i.e. low-temperature, Kondo regime (2 K $\approx0.6\times T_{K1}$) but also to escape the spin-glass order regime (2 K $>T_{g}$). However spin-glass transition effects do affect the magnetization in moderate and large fields, even well above $T_g$ (See for 
instance Ref. \onlinecite{Trans-SG}). It is because the terms non-linear in the field of the magnetization increase rapidly (up to  exhibit a divergence) when $T$ is decreased  down to $T_g$. Thus, to avoid this additional difficulty, we restricted the magnetization analysis to the samples of $T_{g}$ smaller than about one tenth of the measurement temperature 2 K. This implies to restrict the study to our less magnetic samples, the more magnetic one among them being $C-a$ of $T_{g}$= 230 mK. The $M(H)$ data of some of these 
samples are shown in Fig.\ref{figM1}. To analyze our data, we intend to use formula Eq.\ref{eqM1}. But, in this formula, the parameter $x_2$ appears explicitly whereas we only deduced 
previously the value of $x_{2}^{*}$ ($=q\times x_{2}$) for each sample. As the $q$ value is unknown, we transform Eq.\ref{eqM1} as follows:

\begin{figure}
\centering{\includegraphics[width=7cm]{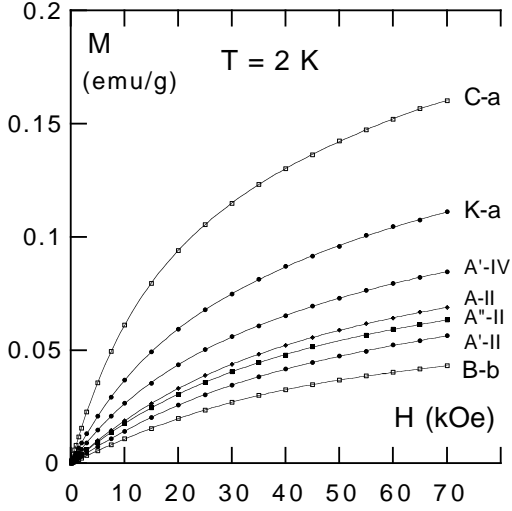}}
\caption{Magnetization at 2 K of some of the less magnetic samples.}
\label{figM1}
\end{figure}

\begin{figure}
\includegraphics[width=9cm]{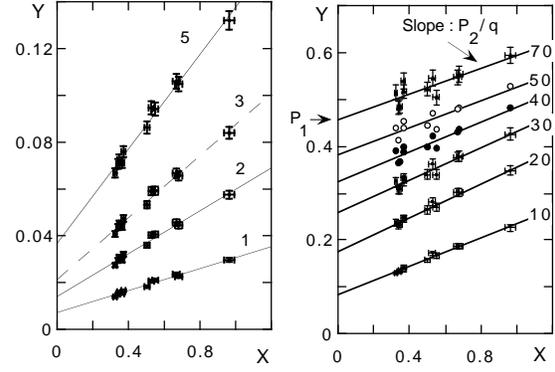}
\caption{Scaled magnetization $Y=M/(Ng\mu_{B} \times x_{1}S)$ at constant magnetic field  vs.
X=$x_{2}^{*}/x_{1}$. Figures are values (in kOe) of the applied
magnetic field. The data $X$ = 0.9 are obtained for the more magnetic
($C-a$) of the samples analyzed here. Error bars, represented only for several constant field values, come from relative errors, 3\% at maximum, for the $x_{1}$ and $x_{2}^{*}$ deduced from the $\chi(T)$ analysis} 
\label{figM2}
\end{figure}
\begin{figure}
\includegraphics[width=8cm]{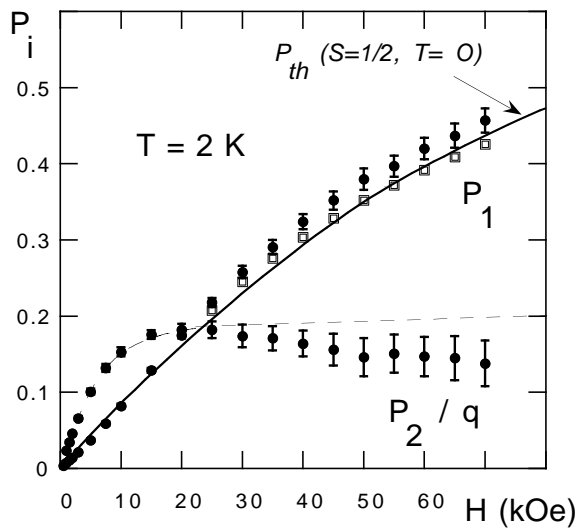}
\caption{Field dependence of $P_{1}$ and $P_{2}/q$ (full
circles) deduced from the $Y(X)$ linear fits of Fig.\ref{figM2}. The error bars originate 
in the scattering of the $Y$ and $X$ data. The open squares represent the $P_{1}(H)$ calculated when a constant value
($\approx 0.2$) of $P_{2}/q$ is imposed in the $Y(X)$ fit at $H>$ 25 kOe. The dashed curve
represent the fit of $P_{2}/q$ by a Brillouin function up to 25 kOe and its
extrapolation up to 70 kOe. The full curve represents the theoretical
variation of the Kondo polarization for spins 1/2 at 0 K (See text).}
\label{figM3}
\end{figure}


\begin{equation}
\  \frac{M(H,T)}{Ng\mu_{B}  x_{1}S} = P_{1} (H,T,T_1^{*})+ \frac{x_{2}^{*}}{
x_{1}} \frac{P_{2}(H,T,T_2^{*})}{q},
\label{eqM}
\end{equation}
The formula Eq.\ref{eqM} is interesting because it contains only two 
unknown parameters: $P_{1}$ and $P_{2}/q$. Indeed, the sample dependent parameter $x_{2}^{*}/x_{1}$, denoted $X$ in the following, can be calculated for each sample from 
our susceptibility results. It increases with the moment concentration since $x_{2}^{*}/x_{1}\propto x_{eff}^{0.55}$ following the results in Sec. \ref{2-TK-X}. The parameter $M(H,T)/ Ng\mu_{B} x_{1}S$, that we denote $Y$, can be deduced from the measured $M(H,T)$ and the $x_{1}$ value. 
The value of $T$ has been fixed: $T=2$ K. Let us fix the value of $H$ and analyze $Y$ as a function of $X$. One finds $Y$ to vary linearly with $X$, whatever the fixed field value: See Fig.\ref{figM2}. This proves that the magnetization obeys Eq.\ref{eqM} with $T_1^{*}$ and $T_2^{*}$  independent of the moment concentration. Therefore, we proved that the simple decomposition of the magnetization in terms of two (and only two) independent contributions holds up to large fields. The next step is to examine the field dependence of $P_{1}$ and $P_{2}/q$ at 2 K. At each field value, the values of $P_{1}$ and $P_{2}/q$ are given by the best linear fit of the $Y(X)$ data. We represent them as a function of the field in Fig.\ref{figM3}. 
In this Figure, one observes readily that $P_{1}(H)$ and 
$P_{2}(H)$ behave in a completely different way. $P_{1}$ increases 
continuously with
the field, with no tendency to reach its saturated value, even at 70 kOe. This feature is thus in strong contrast with the magnetization of
non-interacting spins 5/2, which reaches almost its saturation value above 10 kOe at 2 K. It suggests that $P_{1}$ is governed by a large 
characteristic temperature over the whole field-range. Even more striking, 
the evolution of $P_{1}(H)$ with $H$ at 2 K (i.e. $\approx 0.6\times T_{K1}$) appears to be similar to that of the theoretical Kondo polarization 
$P_{th}(H/T_{K})$ 
calculated at $T=0$ for spins 
1/2. In Fig.\ref{figM3}, the full curve is that of Figure 6.2 in
Ref.\onlinecite{Hewson} rescaled with our $T_{K1}$ value (3.35 K) \cite{footnote}. This quantitative similarity calls 
for three comments. First, it suggests strongly that $P_{1}$ is governed by only one 
energy scale: $T_{K1}$, over the whole studied field-range. Thus $T_1^{*}=T_{K1}$ in formula Eq.\ref{eqM1} for fields from 0 up to at least 70 kOe ($\mu_B H/k_B T_{K1}$ from 0 to 1.4). Second, because the saturated value of $P_{th}(H/T_{K})$ equals 1 in the calculation, that of $P_1 (H)$ should do the same, which was not among the starting hypothesis. The third comment is about the fact that the $P_{th}$ 
has been calculated for spins 1/2 whereas the $Mn$ spins are expected 
to equal 5/2. In 
Appendix \ref{XK}, we recall the result obtained for the susceptibility in the $(n=2S)$-model: 
A Kondo moment behaves as $n$ Kondo spins $1/2$ in view of formula Eq.\ref{eq6}. One of the best possibilities for checking the 
latter prediction consists in studying the field-curvatures of 
the spin polarization. Presently, the 
close agreement between $P_{1}(H, 2K)$ and $P_{th}(H,T=0)$
suggests strongly that this prediction is correct and indeed valid up to fields at which 
the polarization is rather large: See Fig.\ref{figM3}. However, some details of the $P_{1}$ behavior remain obscure. Up to 3 kOe, $P_{1}$ 
is found to vary linearly with an initial slope identical with the previous $\chi_{K1}$ results at 2 K. Note that at this temperature, $\chi_{K1}$ equals about 80\% of its value at $T=0$, hence $P_{1}(H,T=2K)=0.8\times P_{th}(H,T=0)$ at low fields. But at larger fields, up to 20 kOe, $P_{1}$ varies slightly more rapidly than the field, which makes the agreement between $P_{1}(H, T=2K)$ and $P_{th}(H,T=0)$ closer than it could be expected. This feature cannot be explained by the scattering of the $Y$ and $X$ data and  is not predicted theoretically, at least not for spins 1/2. 
Let us now comment on the $P_{2}/q$ behavior. It appears to reach a 
saturation value very rapidly when compared to the $P_{1}$ case, which tends to prove that the  
characteristic temperature for the interactions is very low 
compared to 2 K. It was the result obtained previously in the zero 
field limit. Thus we propose: $T_2^{*}\approx 0$ in formula Eq.\ref{eqM1} whatever the field value. Note that the 
decrease of $P_{2}/q$ at increasing fields above 30 kOe could be 
nothing else that an artifact due to the errors in the
determination of $P_{1}$ and $P_{2}$. Indeed, the relative weight of the
$P_{2}/q$ contribution with respect to $P_1$ decreases at increasing fields. This is obvious in Fig.\ref{figM2}. In turns, to impose a constant $P_{2}/q$ value equal to 0.2 in
the linear fit of $Y(X)$ above 25 kOe does not change drastically the 
$P_{1}$ values obtained previously: See the open squares in 
Fig.\ref{figM3}. 
In summary, we have found that the split of the low field 
magnetization (susceptibility case) into 
two contributions, each governed by a different energy scale (the $T_{K1}$ and an almost vanishing $T_{K2}$), holds up 
to rather large fields over the whole studied field range. We found the 
$P_{1}$ to be described by the theoretical polarization calculated for spins 1/2. 
Note that we found a saturated value of $P_{2}/q$ nearly equal to 0.2, Then, if one assumes the saturated value of $P_{2}$ to be equal to 1, one deduces a $q$ value equal to about 5. This point is discussed in the next section.

\section{Summary and discussion}
\label{dis-2TK}

Preliminarily, we have recalled the two opposite limiting cases, pure RKKY and pure Kondo, associated with two distinct energy scales: The $T_J$ ($\propto x$) which characterizes pure (no Kondo) RKKY coupling effects and the $T_{K0}$ which characterizes the single impurity Kondo interaction. Our more magnetic sample ($R$ sample) seems to fall in the first limiting case, that of a $x$ sufficiently large so that only RKKY features are detected. When decreasing the moment concentration, we observed Kondo susceptibility features at moderate and large temperatures and a spin glass transition at lower temperature. This led us to analyze the susceptibility above $T_g$ with pure Kondo models, ignoring temporarily the RKKY interactions. The only additional hypothesis, that of a broad $P(T_K)$, was introduced because  we found no saturation of the susceptibility at very low temperature. We found a way for fitting easily our data: It consists in replacing the broad $T_K$-distribution by a bimodal one. Indeed, both give  the same averaged Kondo susceptibility in our $T$-range, as we showed from a very simple calculation. 
Of course, because of the bimodal distribution treatment, the susceptibility takes simply the form of a sum of two contributions. \textit{A priori}, there is no physical meaning in this decomposition which can be achieved for any given sample. It is only by studying the characteristic features of each contribution when varying the moment concentration by two decades that we hit on a remarkable result: 
The behaviors of the two limiting cases (pure Kondo, pure RKKY) appear to coexist in the same sample whatever $x$ and are sufficient to describe the  $\chi (T)$ and the $M(H)$. In particular, this decomposition holds up to the moment concentration where the Kondo susceptibility features disappear completely. Let us summarize. One contribution appears to obey the characteristic $x$-independent features of single Kondo impurities in as much as the temperature dependence of the susceptibility (the $\chi_{K1}$) is concerned as well as the low-$T$ field dependence of the polarization (the $P_1$). Consequently, we conclude that  $T_{K1}$ is the bare Kondo temperature $T_{K0}$. The other contribution is that one expected for a pure RKKY spin-glass: 
The $T_{K}$ is (almost) vanishing, the susceptibility (the $\chi_{K2}$) varies as $1/T$ above $T_{g}$,  
the $T_{g}$ varies with $x_{2}^{*}$ as the $T_g$ of standard spin-glasses does with the moment 
concentration and, finally, well above $T_g$, the spin polarization (the $P_2$) reaches its saturated value in a moderate field (at least at 2 K). At  varying moment concentration, the characteristics of the two contributions remain unchanged, only their relative weights vary, obeying $x_2^{*}\propto x_1^{1.8}$ within our concentration range (Sec. \ref{two-TK effect}). When assuming that $x_{eff}$ does reflect the (\textit{a priori} unknown) moment concentration, we found 
that the purely Kondo contribution ($\propto x_1/x_{eff}$) decreases while the RKKY contribution ($\propto x_2^{*}/x_{eff}$) increases when increasing the moment concentration. 
Note that even for samples where the Kondo couplings are dominating, thus of very small $x_2^{*}/x_{eff}$, the $\chi_{K2}$ contribution ends to be spectacularly dominant at low $T$. Indeed, when decreasing the temperature, the $\chi_{K1}$ tends to reach its saturated value whereas the $\chi_{K2}$ contribution continues to increase. An example is given for sample $B-b$ in Fig.\ref{fig2d}: See the $\chi_{K1}(T)$ and $\chi_{K2}(T)$ evolution with temperature. Of course this $\chi_{K2}$ dominance at low $T$  is more and more evident when the moment concentration, and thus the $x_{2}^{*}$ weight, increases: Compare Fig.\ref{fig7} with lower part of Fig.\ref{fig2d}. This trend is apparent even above 2 K as it can be seen in  Fig.\ref{fig03}. Note that finding $T_{K1}$ (identified to $T_{K0}$) constant whereas the moment concentration $x$ is varied by orders of magnitude proves that the electronic structure is not affected by large $x$ changes. In that sense, our system exhibits the same remarkable property as the heavy fermion $Ce_{1-x}La_xPb_3$ where, fortuitously, $T_K$ and the crystal field splitting are concentration independent, which enabled a proof of single moment properties \cite{Lin-87}. In $Al-Pd-Mn$ QC, this property can be attributed to the fact that the moment concentration remains small ($<1 \%$). More fundamentally, it was expected because large variations of the moment concentration can be achieved with minimal changes of the overall number of $Mn$ atoms. 

Our results are the more striking as they were obtained from an analysis using only very general hypothesis from the outset. At this stage, let us attempt to explain our results with different models. First, the problem can be treated on a macroscopic scale. For instance, one can assume that the thermodynamic quantities, $\chi (T)$, $M(H,T)$ or the magnetic specific heat $C_P(T)$, can be developed as a series of terms reflecting successively  a single-impurity effect (pure Kondo), a two-impurity effect (RKKY)...  in the way described in Ref.\onlinecite{Affleck}. But other hypothesis exist in terms of microscopical models able to yield $T_K$ distributions. In such models, broadly distributed environments of the spins are required to generate a $P(T_K)$. Let us consider several models. The first one is based on strong fluctuations of the local DOS $\rho_{local}$ assumed to trigger the $T_K$ value\cite{Kondo desordre} (as well as the RKKY couplings\cite{Mayou}) instead of the average DOS. To apply this model in case of QC is conceivable due to the multiplicity of local \textit{atomic} environments and specific electronic structure\cite{Sig-pre}. However this model can hardly provide an evolution of the $\rho_{local}$ distribution with the moment concentration able to account accurately for the evolution of the $x_i$. Let us now consider microscopical models based on magnetic proximity effects assumed to decrease locally $T_K$ with respect to $T_{K0}$. Then the $T_K$ distribution is due to the multiplicity of local \textit{magnetic} environments (distances between spins, local spin concentration). This model implies a $x$-dependent $P(T_K)$. The upper limit of $P(T_K)$ equals $T_{K0}$. The broadness of  $P(T_K)$ is achieved for a lower limit of $P(T_K)$ (almost) vanishing when compared to $T_{K0}$. In the general case, such a $x$-dependent $P(T_K)$, even with the upper bounding limit $T_{K0}$, cannot be expected to give a $x$-independent $T_{K1}$ when interpreted in terms of a bimodal distribution. 

However, it is the latter model that we develop in the following because it can be easily transformed into an \textit{ad hoc} model able to account for our results. Indeed, one can introduce the following two additional assumptions which forces the $T_K$-distribution to be really bimodal. The first one is a single-impurity Kondo behavior, with $T_K= T_{K0}$, robust enough to apply as soon as the spins are separated by a distance larger than a characteristic length $R_C$. The second (and perhaps stronger) hypothesis is a $T_K$ collapsing very rapidly as soon as the distance between spins becomes smaller than $R_C$. In summary, a spin ${\bf S_i}$ can be considered as magnetically 'isolated' with a $T_K$ equal to $T_{K0}$ if no other spin lies in a sphere of radius $R_C$ centered at site $i$. In the opposite case, spin ${\bf S_i}$ is 'non isolated' and its  $T_K$ is assumed to be (almost) vanishing. In our case, the $T_{K1}$ could characterize the behavior of 'isolated' spins and thus could be identified with $T_{K0}$. 
Let us focus on $R_C$. It may refer to the length scale introduced in different scenario for the $T_K$-collapse due to proximity effects: See the discussion in Ref.\onlinecite{Affleck}. The first one is about Kondo cloud overlap:  The conduction electrons involved in the different Kondo clouds of radius $\xi_K$ are assumed to be almost orthogonal. So, the single-impurity Kondo model should hold even when $\xi_K$ is larger than the distance between moments. Its breakdown has been predicted only for distances between moments much shorter than $\xi_K$, at least at 3D\cite{Affleck}. This leads to a $R_C$ which thus is not too large although the formula for $\xi_K$ ($\xi_K= \hbar v_F/k_B T_K $) yields a huge $\xi_K$ value, lying in the ${\rm  \mu m}$ range for $T_K \sim 1{\rm K}$ and if one identifies $v_F$ with the standard Fermi velocity of pure simple metals. However, the latter assumption is controversial, the $\xi_K$ could be much smaller. Moreover in QC, the Fermi velocity as defined for free electrons is not pertinent because of the specific electronic properties of QC (flat bands)\cite{Mayou-2}. Another possible mechanism for the local $T_K$ depression is a RKKY coupling directly competing with the Kondo interaction. Then, one has to compare $T_{K0}$ with a \textit{local} energy scale $T_{J.loc}(i)$ which reflects the RKKY interactions that spin ${\bf S_i}$ undergoes from the other spins. 
One can write: $T_{J.loc}(i) =   | {\bf S_i}\sum_{j}  J_{ij}{\bf S_j} |  $ from which a local length scale $R_{i}$ can be defined: $1/R_{i} ^{3} \propto | \sum_{j} ({\bf S_iS_j}/S^{2}) \times cos (2kr_{ij} )/r_{ij} ^{3} | $ so that $T_{J.loc}(i) \propto S^{2}/ R_{i} ^{3}$. Then a $R_C$ can be naturally defined as the length for which $T_{J.loc}(R_C) \sim T_{K0}$. Consider the two following limiting cases when the temperature decreases. First, if $T_{K0}\gg T_{J.loc}(i)$ ($R_{i}\gg R_C$), the individual moments are Kondo quenched before the RKKY coupling can become effective. Then the $T_K$ of spin ${\bf S_i}$ equals $T_{K0}$. Second, for $T_{J.loc}(i)\gg T_{K0}$ ($R_{i}\ll R_C$), the RKKY coupling is fully effective well above $T_{K0}$, impeding the ${\bf S_i}$ to be quenched at lower temperature \cite{Schlottmann-98}. In the latter case, when only two spins are involved ($T_{J.loc}(i)= | J_{ij}{\bf S_i}{\bf S_j}  | $), the spins of the two impurities compensate each other for an antiferromagnetic RKKY coupling. But if the RKKY coupling is ferromagnetic, one gets a resulting spin $2S$ of behavior different from that of  single spins\cite{Schlottmann-98}. The problem of comparable $T_{J.loc}(i)$ and $T_{K0}$  is difficult to solve even in the two-impurity case: See Refs \onlinecite{Schlottmann-98} and \onlinecite{Garst-04} and references therein.

In short, we can use a microscopic model where $x_1$ and $x_2$ are actual concentrations of spins, the 'isolated' ones with $T_{K1}=T_{K0}$ and the 'non isolated' ones with $T_K\approx 0$ respectively. Then the actual total moment concentration $x$ equals $x_1+x_2$. Here we note that 
the exhaustion problem is not solved when considering only the 'isolated' spins, although $x_1<x_{eff}$ and $T_{K1}$ is larger than the $T_{Keff}$ considered in Sec. \ref{T>2K}. A calculation similar to that done in Sec.  \ref{T>2K} shows that the number $n_e$ of conduction electrons available for the Kondo screening is, depending on $x_1$, 20 to 300 times smaller than that, $2SNx_1$, necessary to screen all the 'isolated' $Mn$ spins. In addition $T_{K1}$ does not follow qualitatively the $1/x_1$ law predicted to reconcile the exhaustion problem with the possibility to observe a Kondo effect (Sec. \ref{T>2K}). In strong contrast, $T_{K1}$ is constant over our whole concentration range. This calls for further theoretical explanations\cite{Noz-2005}. Granted that, the present microscopic model well accounts qualitatively for our main results, independently of the $T_K$-collapse mechanism. Let us start from extremely low moment concentration: The probability $x_1/x$ to have 'isolated' spins  should be maximum, thus equal to 1. But it is expected to decrease with increasing moment concentration while that ($x_2/x$) of the 'non isolated' spins increases, which agrees with the results shown in Fig.\ref{Ai-norm-xeff} when $x_{eff}$ is identified with $x$. Interestingly enough in this model, when the moment concentration is increased up to the limit where the mean distance $\bar{r}$ between moments equals about $R_C$, only a few spins are still 'isolated'. Then the system should be (almost) fully dominated by the RKKY interactions and the $\chi (1/T)$ curvatures should have vanished. This is precisely what we observed when the moment concentration $x$ is increased up to that of sample $R$. Hence we identify  the value of $R_C$ with that of $\bar{r}$ ($=a\times x^{1/3}$) of sample $R$. Here, $a$ is the inter-atomic distance. If $x$ of sample $R$ is simply deduced from the measured Curie constant (assuming $S=5/2$), one finds a $\bar{r}$ and thus a $R_C$ equal to about $5.3\times a$ ($13\, \AA$). This model also predicts spin-glass transition features which well agree with our results. Indeed, the 'non isolated' spins can be assumed to undergo RKKY interactions since their $T_K$ is vanishing whatever the $T_K$-collapse origin is. In addition, because $T_g$ is smaller than the $T_{K1}$ of the 'isolated' spins, the latter should be Kondo quenched and thus transparent for the RKKY interactions at $T_g$. Hence, for the spin-glass transition, the sample behaves as if the only existing spins are the 'non isolated' ones. It follows a $T_g $ expected to obey a power of $x_2$ with a standard exponent value ($\approx 0.7$), which we precisely found experimentally. 

More surprisingly, in view of the crudeness of the underlying hypothesis, the predictions of the present model account even quantitatively, and not only qualitatively, for our results, as shown below. This is the more correct, the smaller the moment concentration in order to have only 'isolated' spins and pairs, but not triplets, of spins with $r_{ij}<R_C$. Then, the pair concentration is small, so each pair, which occupies a small region (linear size $R_C$) in the space, is separated from the other ones by large distances. Consequently, the RKKY interaction should be rather strong between two spins of a pair (because $J_{ij}\propto 1/r_{ij}^{3}$ and $r_{ij}<R_C$), but much smaller  between the spins of two different pairs because of the distance ($\gg R_C$) between pairs. Then one expects the RKKY coupling to generate distinct effects depending on  the temperature range. 
When the temperature is decreased, the first effective RKKY interactions are those coupling two spins in a pair, forming a ferromagnetic $F$-pair (spin $2S$) or an antiferromagnetic $AF$-pair (spin 0) depending on the value of $k_F r_{ij}$. It is only at much lower temperature that the spins $2S$ of the $F$-pair undergo RKKY couplings from the other ones to yield a spin-glass transition. This picture implies that $P_2$, which becomes a polarization of true moments in the model, is a polarization of $F$-pairs  at moderate $T$. This agrees with our $P_2/q$ results for the less magnetic samples (Sec. \ref{Mag-2K}): We found $P_2/q$ at 2 K to scale with a Brillouin law (See Fig.\ref{figM3}) with a spin value equal to nearly 4, thus comparable with the resulting spin of a $F$-pair of two spins $S=2$. 
However, two additional conditions have to be introduced to achieve a quantitative agreement. The first one is an equal probability to have $F$ and $AF$ pairs, which is expected for perfect disordered configurations because of the very rapid oscillations of the RKKY coupling with distance within $R_C$. The second condition is a large $S$ value. This allows to account for $\chi_{K2}$ varying simply with $1/T$ over the whole $T$-range above $T_g$ because the Curie constant for a collection of such pairs can be considered as temperature independent. As an example, consider only 4 individual non Kondo spins $S$: Their Curie constant equals $\alpha \times 4\times S(S+1)$ while that of two pairs of spins, one $F$-like (spin $2S$), the other $AF$-like (spin $0$), equals $\alpha \times 2S(2S+1)=\alpha \times 4\times S(S+1/2)$. Hence the Curie constant is almost the same for large $S$ values, whether the spins are RKKY-correlated in pairs at low temperature or uncorrelated at large temperature. A straightforward calculation provides a similar result in case of triplets, quadruplets ... when the ferromagnetic and antiferromagnetic RKKY couplings are of equal probability. In our case, if we assume $T_{K2}=0$, we can write $\chi _{K2}=\alpha \times x_2 \times Sq/T$ with $q$ varying from $S+1/2$ to $S+1$ when increasing the temperature. In contrast, the present pair hypothesis modify the magnetization analysis. Indeed, when the moments of each pair are correlated, only $F$-pairs (spin $2S$, concentration $x_2/4$) contribute to the magnetization: The latter equals $Ng\mu_{B}\times (x_{2}/4)\times 2S 
\times P_{2}$, i.e. half the second term of $M$ in Eq.\ref{eqM1}. This implies a modification of formula Eq.\ref{eqM} as follows: $Y=P_{1}+X\times P_{2}/2q$. Hence it is the saturated value of $P_{2}/2q$, instead of that of  $P_{2}/q$, which has to be identified with the value 0.2 in Sec. \ref{Mag-2K}. It follows $q=2.5$ since $P_2$ is a normalized polarization in the model, thus of saturated value equal to 1. Thus, we obtain three converging results when comparing the model and the data and assuming $T_{K2}$ to equal strictly zero: $S=2$ from the Brillouin fit of $P_2$, $q=(S+1/2)$ to $(S+1)$ from the susceptibility and $q=2.5$ from the saturated value of the magnetization. So, at least for low moment concentrations, the present microscopic model appears to well account for the RKKY dominated term of the magnetization provided that one assumes $T_{K2}=0$, an effective $Mn$ spin equal to 2 and an equal number of the ferromagnetic and antiferromagnetic pairs.
\begin{figure}
\includegraphics[width=8cm]{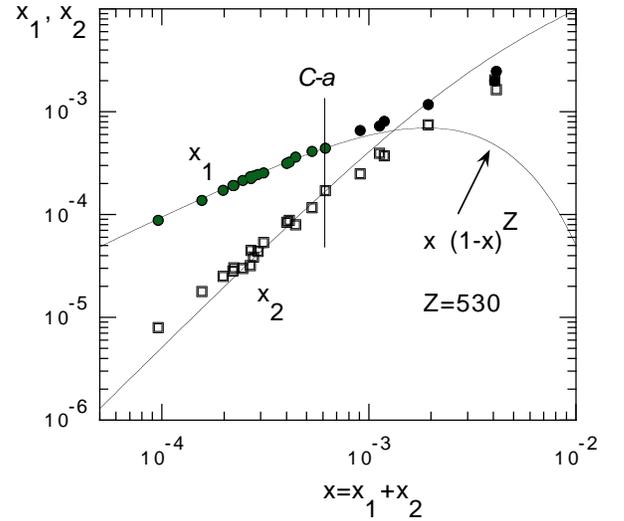}
\caption{$x_{1}$ (solid circles) and $x_{2}=x_{2}^{*}/q$ (open squares) are plotted vs. $x_{1}+x_{2}$ in a Log-Log diagram. Here, $q=2.5$. The solid curve are the fitting curves with a simple $x_{i}(x,Z)$ dependence (See text), which roughly works only for samples less magnetic than sample $C-a$. }
\label{fig.end}
\end{figure}

The present model also allows to readily express the concentrations $x_1$ and $x_2$ in terms of $\pi_0$ the probability for a spin to be 'isolated' as previously done in Refs \onlinecite{Tholence} and \onlinecite{Tissier}. In case of a perfect random distribution of the moments in space, $1-x$ is the probability for any given atomic site to carry no spin. Hence $x_{1}=x\times {\bf \pi_0}$ and $x_2=x\times (1-{\bf \pi_0})$ where ${\bf \pi_0}= (1-x)^{Z}$ with $Z$ the number of atomic sites in a sphere with radius $R_C$: $Z=(4\pi/3) \times (R_C/a)^{3}$. 
For vanishing $x$ ($x\ll 1/Z$), $x_{1}\lesssim x$ and $x_{2}\approx Z \times x^{2}$ (thus $\approx Z \times x_1^{2}$). The above (non simplified) $x_{1}(x,Z)$ and $x_{2} (x,Z)$ laws does not allow to fit, even poorly, our data over the whole $x=x_1+x_2$ range. Here we calculate  $x_{2}$ using $x_{2}=x_{2}^{*}/q$ and $q=2.5$ from the previous discussion. However the fits can be forced for samples as or less 
magnetic than $C-a$ ($x=6\times10^{-4}$): See Fig.\ref{fig.end}, providing a value for $Z$: $ \approx 530$. (Note that we find actually $x_{2}\propto x_1^{1.8}$ and that over our whole concentration range, see Sec. \ref{two-TK effect}.) Strikingly, this $Z$ value yields a $R_C$ value ($\approx 5\times a$), thus obtained only for weakly magnetic samples, in excellent agreement with that ($5.3\times a$) obtained from the above arguments on the disappearance of the Kondo $\chi (1/T)$ curvatures in  the more magnetic samples. 

Up to now, we did not find any difference between the predictions of the Kondo cloud overlap and local competition RKKY$\times$ Kondo pictures, especially for low moment concentrations. However, the latter model could better explain the following two features exhibited by the more magnetic samples. First, for a sample of mean distance $\bar{r}$ between moments comparing with $R_C$, most $T_{J.loc}(i)$ compare with $T_{K1}$. Now the value of $T_g$ reflects that of the $T_{J.loc}(i) $ averaged over all spins when no spin is Kondo-like. It follows that the $T_g$ for a sample of $\bar{r}\sim R_C$ should compare with $T_{K1}$ in the present assumption. It is exactly what we found experimentally for sample $R$ where we associated $\bar{r}$ with $R_C$: Its $T_{g}$ (3.6 K) compares with $T_{K1}$ (3.35 K). The second feature is the discrepancy found between the experimental $x_1$ and $x_2$ and the $x_1 (x,Z)$ and $x_2 (x,Z)$ calculated for a perfect random distribution of $Mn$ spins in the space. This discrepancy is all the more marked when the moment concentration increases: See the data and the $x_1 (x,Z)$ and $x_2 (x,Z)$ fitting curves extrapolated from the low moment concentration regime in Fig.\ref{fig.end}. A local RKKY $\times$ Kondo competition assumption allows to reconcile the data with the predictions made in case of a purely random spatial distribution of the moments. Indeed, a spin $S_{j}$ separated from spin $S_{i}$ by a distance smaller than $R_{C}$ can generate a RKKY effect at site $i$ partially cancelled by other spins lying more or less far away from site $i$. In consequence, a spin $S_i$ at the center of a sphere of $Z$ sites ($Z\propto R_{C}^{3}$) can be 'isolated' although other spins may lie inside the sphere. It follows that $x_{1}$ could be larger than $x\times (1-x)^{Z}$ at large concentration. In conclusion, microscopic models, especially that of a local RKKY$\times$ Kondo competition, account quantitatively for our results, including many details, over the whole here-studied [$T,H$] range. 

Let us end the present discussion by comparing our method of analysis and our results with previously published works where thermodynamic quantities (susceptibility, specific heat) were decomposed into two contributions. First, let us quote the magnetization analysis performed for dilute alloys, first for $Fe$ in $Cu$ in the early 70's\cite{Tholence}. The authors built a microscopic model of 'isolated' Kondo impurities and (almost non Kondo) impurity pairs to account for their results. In this case, because of the solubility limit of $Fe$ in $Cu$, the largest studied moment concentration was small, equal to that of our sample $C-a$ ($ 6 \times 10^{-4}$). Interesting enough, our results are similar to those for $Fe$ in $Cu$ in the same $x$-range ($< 6 \times 10^{-4}$). But, in addition, we could observe sizable low temperature $P_1(H)$ curvatures up to 70 kOe, which was not possible for $Fe$ in $Cu$ due to its large $T_{K1}$ value ($\sim 12$ K deduced following our procedure). For $Fe$ in $Cu$, one could argue that the magnetization decomposition was actually forced by a chemical segregation of the $Fe$ atoms into pair, triplets ... . But more essentially, the smallness of the $Fe$ concentration, and thus of the averaged RKKY couplings, made these samples mainly Kondo dominated. Then one could argue the RKKY effect to give only a perturbative correction which simply adds to the Kondo susceptibility. Our results bring the question up again since, actually, the same magnetization decomposition holds up to concentrations where the RKKY couplings become dominant and thus cannot be anymore considered as a perturbation. Another example in the literature is the analysis of the magnetic specific heat $C_P$ in $Ce$ alloys\cite{Sereni}. It is the case for instance of $Ce(In_{3-y}Sn_y)$ where the moment concentration (which is large) is kept constant but where the random $Sn$ substitution to $In$ atoms induces environment fluctuations around each $Ce$ atom. This makes the analysis much more complicated than in our case because the characteristics of each contributions (Kondo, RKKY) and not only their relative weight depend on the composition parameter $y$.

\section{Conclusion}
\label{conclusion}

We studied a system (QC $Al-Pd-Mn$) which belongs to the same category as the dilute alloys concerning the problem of the RKKY$\times$Kondo competition. However, it is a case where the few $Mn$ atoms which carry an effective spin can be of broadly varying  concentration whereas the $Mn$ atoms are a component of the structure, of nearly fixed concentration. Because of the magnitude of the involved characteristic energy scales and the possibility to tune the moment concentration up to a point where the Kondo features disappear completely, we could obtain relevant results that we analyzed with no preconceived assumptions. Surpringly, we were led straightforwardly to a very simple result: The magnetization (and thus the susceptibility) is the sum of two contributions, one purely Kondo ($T_{K1}=3.35$ K), the other one dominated by RKKY couplings. These two contributions coexist with the same characteristic temperatures, whatever the moment concentration up to that one where the RKKY interactions
dominate fully. When building microscopical pictures, we found that different mechanisms (Kondo cloud overlap, RKKY) for breaking locally the Kondo effect account quantitatively for our data with an equal success. It is only by an analysis of the more magnetic samples, where the Kondo effects become negligible, that we could propose the local RKKY$\times$ Kondo competition as the best candidate to  explain the local $T_{K}$ depression. 
To conclude, we can ask the question whether all these models really correspond to a physics relevant at a microscopic level or provide a predictive picture for actually a more macroscopic and subtle physics such as two-fluids models do.

\appendix

\section{Single-impurity Kondo results}
\label{XK}

Only one energy scale, the Kondo temperature $T_{K}$, accounts for the single impurity magnetic behavior in the pure spin case (orbital moment $L= 0$). Hence a Kondo susceptibility $\chi_K (S,T,T_K)$ depending only on $T$,  $T_K$ and the spin value $S$. The Bethe $ansatz$ calculations performed in the frame of the ($n=2S$)-model showed that the temperature dependence of the susceptibility is the same for a spin 1/2 and a spin larger than 1/2 \cite{Desgranges,Sacramento}. So we first recall the results for a spin 1/2. In the following, formula are written for a collection of spins, with concentration $x$, and $N$ denotes the total number of atoms. 
Wilson \cite{Wilson-1} showed that the susceptibility $\chi_{K}$ can be expressed in
terms of an implicit equation of the form: $\phi (y) = {\rm ln} (t)$ 
where $\phi(y)$ is a universal function of $y$. Here, $t=T/T_{K}$, $y$ is the relative deviation of the Kondo susceptibility $\chi_{K}$ from the Curie law $C/T$: $y= (\chi_{K}-C/T)/(C/T)$ and the Curie constant $C$ is calculated for a spin 1/2: $C= \alpha   x  (3/4)$ with $\alpha=N(g\mu_B)^{2}/3k_B$. To evaluate $\chi_K (T,T_K)$ amounts to calculate $y(t)$. One can write $\chi_{K}(T,T_K)=(C/T_K) (y(t)+1) / t$ and thus $\chi_{K}(T) = \chi_{K} (T=0) F(t)$ where $\chi_{K} (T=0)= Cw/T_{K}$ and, by definition, $F(t=0)=1$. The value of the parameter  $w$ depends on the definition of $T_{K}$. That given by Wilson from a precisely defined procedure extended to very high $T$ \cite {Wilson-1} yields $w = \pi^{-3/2} \times e^{(c+1/4)}$ = 0.4107\ldots ($c$ is the Euler's
constant) \cite{Hewson}. In summary:

\begin{equation}
\ \chi_{K}(S=1/2,T,T_K) = \frac{0.308 \alpha x}{T_{K}}  F(\frac{T}{T_{K}}),
\label{eq4}
\end{equation}

We used the following results in order to build the function $F(T/T_{K})$. For $T/T_{K}>30$, $y$ remains 
of moderate value and perturbation theory can be applied. Hence we can deduce $\chi_{K}$ (to better
than 1\%) from an 
asymptotic analytic formula for $\phi (y)$. Following Wilson\cite{Wilson-1}: $\phi (y)
= -1/y - 0.5 {\rm ln}\vert y\vert +1.5824 y$. 
For lower $t$ values, the behavior is fundamentally 
non-perturbative. Thus, for $t$ = 0.1 to 28, we used the
renormalization results in Ref.\onlinecite {Wilson-2}. Close to $T=0$, the
Fermi liquid approach \cite {Noziere} provides: $\chi_{K}(T)/\chi_{K} (T=0)= 1 - a t^{2}$ where
$a= \sqrt{3} \pi^{3}w^{2}/8$ (half the value of the $a$ in Ref.\onlinecite {Hewson} due to a typing error in the quoted book). Note that one can approximate the Kondo susceptibility by a Curie-like term at large temperatures. It is because, the $\chi_K(T)$ is dominated by the $C/T$ term and  the corrections, because they are logarithmic, can be averaged over a 
restricted $T$-range. For instance, $F(t)/T_{K}$ in formula Eq.\ref{eq4} nearly equals $ 1.65/T$ for $20<t< 1.5\times10^{3}$, hence: $\chi_{K}=C'/T$, with $C'=1.65\times Cw\approx 0.68 \times C$, to better than 2\%. 

To treat the case $S>1/2$, one can use the $n$-channel Kondo model with $n=2S$ which yields a completely spin-compensated ground state. In the frame of this model, Bethe $ansatz$ calculations which previously were shown to treat successfully the $(S=1/2)$-case\cite{andrei-1983} have been carried out for $S>1/2$. The results are 
the following\cite{Desgranges}. For $n=2S$, the susceptibility reaches a saturated value $\chi_n(T=0)$ at $T=0$. When one defines the Kondo temperature $T_K'$ as given from formula\cite{Hewson,Desgranges,Sacramento}: $\chi_n(T=0)\propto n /\pi T_K'$ (hence $\propto S /\pi T_K'$ and not $\propto S(S+1) /T_K'$), $\chi_n(T)/\chi_n(T=0)$ is found to be an almost $n$-independent function of $T/T_K'$ at low and moderate $T/T_K'$. Hence the ($n=2S$)-expression : 
\begin{equation}
\ \chi_{K}(S,T,T_K') = \frac{3\alpha x S}{\pi T_{K}'}  \psi (\frac{T}{T_{K}'}) \hspace{0.2cm} {\rm with} \hspace{0.2cm}  \psi (0)=1
\label{eq5}
\end{equation}
where $\psi (T/T_{K}')$ is $S$ independent. Other authors checked the validity of this model by comparing their theoretical results to experimental susceptibility,
resistivity and specific heat data obtained for some dilute
alloys\cite{Sacramento,Schlottmann-89}. Formula Eq.\ref{eq5} is equivalent to: $\chi_{K}(S,T,T_K')=n\times \chi_{K}(S=1/2,T,T_K')$. So, the present definition of $T_K'$ makes the low temperature susceptibility of a spin $S=n/2$ to behave as that of $n$ independent spins 1/2 having the same $T_K'$ value. When identifying the $\chi_n(T=0)$ in Formula Eq.\ref{eq5} for $S=1/2$ with the $\chi_K(T=0)$ of Wilson, one finds the relation between the values of $T_K'$ and the Wilson's $T_K$: $T_{K}'= 2T_{K}/(\pi w)$, which allows to rescale formula Eq.\ref{eq5} using the Wilson's $T_K$:

\begin{equation}
\ \chi_{K}(S,T,T_K) = n \times \chi_{K}(S=1/2,T,T_K),\,\, n=2S
\label{eq6}
\end{equation}
where $\chi_{K}(S=1/2)$ is given in Eq. \ref {eq4}, 
For $T>T_K'$, deviations of $\psi (T/T_K')$ with the $S=1/2$ results occur, increasing with Log($T/T_K'$). These deviations are unnoticeable in the $\chi_n(T)/\chi_n(0)$ curves shown in Fig.1 in Ref.\onlinecite{Desgranges}. But they can be calculated from the $T \chi_n(T)/B_n$ curves in the same diagram. $B_n$ denotes the standard Curie constant:  $B_n\propto S(S+1)$. These deviations are expected since Eq.\ref{eq6} gives an odd limit for $\chi_{K}(S,T,T_K)$, $S>1/2$,  when $T/T_K' \rightarrow  \infty$. Indeed, the expected limit is the susceptibility of non Kondo spins: $\chi= \alpha x S(S+1)/T$ whereas that one calculated from Eq.\ref{eq6} equals $n$ times the limit for $S=1/2$ and thus $\alpha x S (3/2)/T$. From the $T \chi_n(T)/B_n$ curves in Fig.1, Ref.\onlinecite{Desgranges}, one finds a good approximation to account for the present deviations for $1<T/T_K'<20$: $\psi (S=5/2, T/T_{K}')\approx \psi (S=1/2,T/T_{K}')\times (1+\epsilon)$ where $\epsilon=0.0035\times (T/T_K')$. In our case, we found $T_{K1}=3.35$ K ($T_K'=$5.2 K) from the two-$T_K$ analysis carried out for 2 K $\leq T \leq$ 100 K ($0.4<T/T_K'<19$) neglecting these deviations. When using the ($1+\epsilon$)-correction, we find a $T_{K1}$ only about $10\% $ smaller than the previous estimate. We conclude that our analysis above 2 K gives correct results. For $T/T_K'$ much larger than 20, the deviations of $\psi (T/T_K')$ from the $S=1/2$ results are larger. We do not want to make any assumption, except that $\chi_{K}$ should equal $\alpha x S(S+1)/T$ with logarithmic corrections\cite{Andrei-2,Schlottmann-2}. This is the reason for which we write $\chi_{K}=\alpha \times Sq/T$, $q<S+1$, over restricted high-$T$ ranges ($q=S+1$ is achieved for $T_K$ strictly equal to zero in case of finite $T$). Finally, let us give the reason why we chose the $n=2S$
results instead of a Curie-Weiss (C.W.) law: $\chi_{K}= A/(T+\theta)$ to fit
our data. The C.W. fits provide unreliable
estimates of $T_{K}$ since $\theta / T_{K}$ depends on the $t$ range: For
instance, one finds $\theta = 1.34\times T_{K}$ for $1<t<20$, $\theta =
2.77\times T_{K}$ for $10<t<100$, $\theta = 3.32\times T_{K}$ for $20<t<1500$
from the fit of the function $F(t)$ with a (C.W.) law. Therefore, from the $\theta$ value found for
temperatures covering one or two decades of $t$, it is impossible to predict accurately the
susceptibility for much lower (or larger) temperatures.

\section{P($T_{K}$) distribution effect}
\label{P(Tk)}

\begin{figure}
\includegraphics[width=8cm]{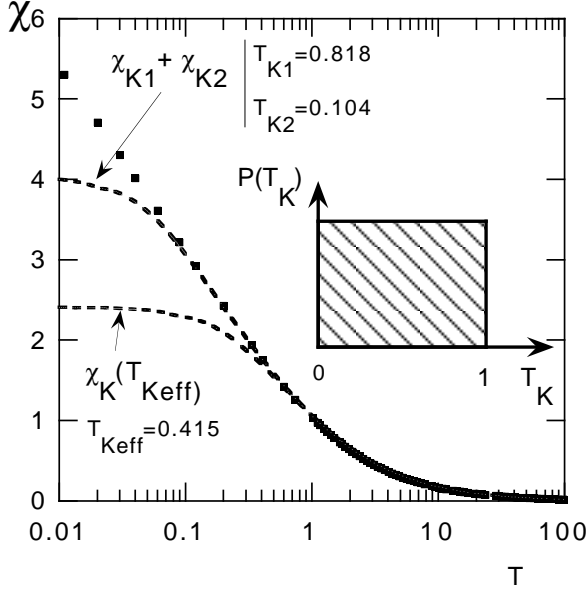}
\caption{The solid squares represent the $\overline{\chi (T)}= \int \chi (T,T_{K})\times P(T_{K}) dT_{K}$ calculated for the $P(T_{K})$ shown in the figure. Here the temperatures are in $T_{Kmax}$ units. The fits with a single-$T_K $ and a two-$T_K$ model for $1<T<100$ are extrapolated down to $T=0.01$, which shows the domain of validity for the approximation of $\overline{\chi (T)}$ as a sum of a limited Kondo terms }
\label{P(TK)}
\end{figure}

In many experimental works devoted to Kondo alloys, the susceptibility has been analyzed with the minimum number of values of $T_{K}$ (or $\theta$ from C.W.
laws), necessary to account for the data. For instance, in the case of dilute alloys such as Cu-$Fe$ \cite{Tholence},
Pt-$Co$ \cite{Tissier}, two $\theta$ values were necessary for an
analysis carried out down to low temperatures. Of course, it is the only
possible method for analyzing the data in a simple way. However, such an analysis can hide the existence of a distribution of $T_{K}$ as we discuss now. Let us
take an over-simplified example, that of a flat normalized distribution: P($T_{K}$)
= $1/(T_{Kmax}-T_{Kmin})$ for $T_{Kmin} \leq T_{K} \leq T_{Kmax}$ and
P($T_{K}$) = 0 elsewhere. The calculation of the average susceptibility
$\overline{\chi (T)}= \int \chi (T,T_{K})\times P(T_{K}) dT_{K}$ is very simple
when one uses a Curie-Weiss-like laws for $\chi (T,T_{K})$: This 
yields qualitatively the results given in the following. However, we used the
Wilson's law \ref{eq4}, namely: $\chi \propto (x/T_{K}) \times F(T/T_{K})$, to calculate
numerically $\overline{\chi}$ in the two following cases. The first case is
that of a $T_{K}$ distribution rather narrow in view of the ${\rm ln} (T)$
susceptibility feature. Taking $T_{Kmin} = T_{Kmax}/3$, one finds a
$\overline{\chi}$ obeying a single-$T_{K}$ law: $\overline{\chi}=
(x^{*}/T^{*}_{K}) \times F(T/T^{*}_{K})$ over a wide T/$T^{*}_{K}$ range, from
$10^{-2}$ to $10^{2}$. The value of $x^{*}$ is found to be almost equal to that
of $x$: $x^{*} =0.96\times x$ and $T^{*}_{K}\approx T_{Kmax}/\sqrt{3}$. This result shows that a narrow distribution of $T_K$ is undetectable from the analysis of measured susceptibility. 
The second case is that of a broad distribution $P(T_{K})$ spreading down to $T_{Kmin}$ = 0. See Fig.\ref{P(TK)}: $\overline{\chi}$ can be accurately fitted over two temperature decades, from
$T/T_{Kmax}$ =1 to 100, by a single-$T_{K}$ Kondo susceptibility with
$T^{*}_{K}=0.415\times T_{Kmax}$. Here again, $x^{*}$ almost equals $x$: $x^{*}
=0.99\times x$. In the same $T/T_{Kmax}$ range, $\overline{\chi}$ can be also
fitted by a two-$T_{K}$ model, which yields: $x_{1}=0.63 \times x$,
$T_{K1}=0.818\times T_{Kmax}$, $x_{2}=0.42 \times x$ and $T_{K2}=0.104\times
T_{Kmax}$. When the $T/T_{Kmax}$-range is extended to a third temperature decade, namely down to 0.05, the $\overline{\chi}$ behavior cannot be described by a single-$T_K$ model anymore, but can be accounted for by two-$T_K$ model with parameters values ($x_{i}$ and $T_{Ki}$)
close to those found at $T/T_{Kmax}>1$. In summary one has: $T_{K2}<  T^{*}_{K} <T_{K1}$ and $x_1+x_2$ close to the $x^{*}$ value found from a single-$T_{K}$ fit performed over the two previous temperature decades. 
Thus, we conclude that, when a sum of
two Kondo susceptibilities is necessary to account for experimental data over a
wide temperature range, one may actually deal with a distribution of $T_{K}$,
spreading over a large $T_{K}$ range.


\begin{thebibliography}{99}


\bibitem{Tholence} J.L. Tholence and R. Tournier, Phys. Rev. Lett. {\bf
25}, 867 (1970)
\bibitem{Hewson} A.C. Hewson, {\it Cambridge Studies in Magnetism},
edited by D.~Edwards and D.~Melville (Cambridge University Press, 1993). 
\bibitem{Noz-Blandin} P. Nozi\`eres and A. Blandin, J. Phys. (Paris) {\bf
41}, 193 (1980)
\bibitem{art_lasjaunias} J.C. Lasjaunias, A. Sulpice, N. Keller,
J.J. Pr\'ejean and M. de Boissieu, Phys. Rev. B {\bf 52}, 886 (1995)
\bibitem{art_Hippert-2003} F. Hippert, M. Audier, J.J. Pr\'ejean, A.
Sulpice, E. Lhotel, V. Simonet and Y. Calvayrac, Phys. Rev. B {\bf 68}, 134402
(2003).
\bibitem{Binder} K. Binder, Solid State Comm. {\bf 42}, 377 (1982).
\bibitem{Sig-pre} J.J. Pr\'ejean, C. Berger, A. Sulpice and Y. Calvayrac,
Phys. Rev. B {\bf 65}, 140203 (2002).
\bibitem{Trans-SG} C. Berger and  J.J. Pr\'ejean, Phys. Rev. Lett. {\bf
64}, 1769 (1990).
\bibitem{art_nmr2} J. Dolin\u sek, M. Klanj\u sek, T. Apih,
J. L. Gavilano, K. Giann\`o, R. H. Ott, J. M. Dubois and K. Urban,
Phys. Rev. B {\bf 64}, 24203 (2001).
\bibitem{art_nmr3} V. Simonet, F. Hippert, C. Berger and Y.
Calvayrac, in {\it  Proceedings of the 8th International
Conference on Quasicrystals, Bangalore, 2002}, J. Non-Cryst Solids 
{\bf 334-335}, 408 (2004).
\bibitem{Desgranges} H.U. Desgranges, J. Phys. C {\bf
18}, 5481 (1985)
\bibitem{Sacramento} P. Schlottmann and P. D.
Sacramento, Adv. in Phys., {\bf 42}, 641 (1993) and References therein.
\bibitem{Wilson-1} K.G. Wilson, Rev. Mod. Phys. {\bf 47}, 773 (1975)
\bibitem{Wilson-2} H.R. Krishna-murthy, J.W. Wilkins and K.G. Wilson, Phys. Rev. B {\bf 21},
1003 (1980).
\bibitem{Noziere-2}  P. Nozi\`eres, Eur. Phys. J B {\bf 6}, 447 (1998).
\bibitem{art_specif1} M.A. Chernikov, A.Bernasconi, C. Beeli, A. 
Schilling and H.R. Ott, Phys. Rev. B {\bf 48},
3058 (1993).
\bibitem{Swenson} C.A. Swenson, I.R. Fisher, N.E. Anderson, Jr., P.C. Canfield and A. Migliori, Phys. Rev. B {\bf 65}, 184206 (2002).
\bibitem{Sacr-1991} P.D. Sacramento and P. Schlottmann, J. Phys.: Condens. Matter {\bf 3}, 9687 (1991).
\bibitem{footnote} In the quoted Figure (where the scaling variable has to be corrected by replacing $H/T_{H}$ by $H/T_{1}$), the author represents the variation of $M(T=0, S=1/2)$ with
$g\mu_{B}H/k_{B}T_{1}$. The $T_{1}$ is defined as follows: $T_{1}= (8/\pi e)^{1/2} \times T_{K}/w$, that we calculated for $T_{K}$=3.35 K to plot the solid curve in Fig.\ref{figM3}.
\bibitem{Lin-87} C.L. Lin, A. Wallash, J.E. Crow, T. Mihalisin and P. Schlottmann, Phys. Rev. Lett. 
{\bf58}, 1232 (1995)
\bibitem{Affleck} V. Barzykin and I. Affleck, Phys. Rev. B {\bf 61},
6170 (2000).
\bibitem{Kondo desordre} V. Dobrosavljevic, T.R. Kirkpatrick and G.
Kotliar, Phys. Rev. Lett. {\bf 69}, 1113 (1992)
\bibitem{Mayou} S. Roche and D. Mayou, Phys. Rev. B {\bf 60}, 322 (1999).
\bibitem{Mayou-2} D. Mayou, Phys. Rev. Lett. {\bf 85}, 1290 (2000).
\bibitem{Schlottmann-98} P. Schlottmann, Phys. Rev. Lett. {\bf 80}, 4975 (1998).
\bibitem{Garst-04} M. Garst, S.Kehrein, T.Pruschke, A. Rosch and M. Vojta, Phys. Rev. B {\bf 69}, 214413 (2004)
\bibitem{Noz-2005} P. Nozi\`eres in \textit{Kondo effect - 40 years after the discovery} \lbrack J. Phys. Soc. Jpn {\bf 74}, 4 (2005)\rbrack
\bibitem{Tissier} B. Tissier and R. Tournier, Solid State Comm. {\bf
11}, 895 (1972)
\bibitem{Sereni} A.M. Lobos, A.A. Aliglia and J.G. Sereni, Eur. Phys. J B {\bf 
41}, 289 (2004).
\bibitem{Noziere} P. Nozi\`eres, J. Low Temp. Phys. {\bf
17}, 31 (1974) and \textit{Low Temperature Physics Conference Proceedings LT14} {\bf
5}, edited by M. Krusius and M. Vioro , (Elsevier, New York, 1975).
\bibitem{andrei-1983} N. Andrei, K. Furuya and J.H. Lowenstein, Rev.
Mod. Phys. {\bf 55}, 331 (1983)
\bibitem{Schlottmann-89} P. Schlottmann, Physics Reports {\bf
181}, 1 (1989) 
\bibitem{Andrei-2} N. Andrei and C. Destri, Phys. Rev.
Lett. {\bf 52}, 364 (1984)
\bibitem{Schlottmann-2} P. Schlottmann and P.D. Sacramento, Physica B
{\bf 206 - 207} 95 (1995)




\end{thebibliography}
\end{document}